\documentclass[sigconf]{acmart}

\usepackage{multirow}
\usepackage{subfigure} 
\usepackage{graphicx}
\usepackage{makecell}
\usepackage{soul}

\AtBeginDocument{%
  \providecommand\BibTeX{{%
    \normalfont B\kern-0.5em{\scshape i\kern-0.25em b}\kern-0.8em\TeX}}}

\copyrightyear{2024} 
\acmYear{2024} 
\setcopyright{rightsretained} 
\acmConference[CHI '24]{Proceedings of the CHI Conference on Human Factors in Computing Systems}{May 11--16, 2024}{Honolulu, HI, USA}
\acmBooktitle{Proceedings of the CHI Conference on Human Factors in Computing Systems (CHI '24), May 11--16, 2024, Honolulu, HI, USA}
\acmDOI{10.1145/3613904.3642859}
\acmISBN{979-8-4007-0330-0/24/05}

\begin{document}

\title[Designing for Human Operations on the Moon]{Designing for Human Operations on the Moon: Challenges and Opportunities of Navigational HUD Interfaces}

\author{Leonie Bensch} 
\affiliation{ 
    \institution{German Aerospace Center (DLR)}
    \city{Braunschweig}
    \country{Germany}} 
\email{leonie.bensch@dlr.de}

\author{Tommy Nilsson} 
\affiliation{ 
    \institution{European Space Agency (ESA)}
    \city{Cologne}
    \country{Germany}} 
\email{tommy.nilsson@esa.int}

\author{Jan Wulkop} 
\affiliation{ 
    \institution{German Aerospace Center (DLR)}
    \city{Braunschweig}
    \country{Germany}} 
\email{jan.wulkop@dlr.de}

\author{Paul de Medeiros} 
\affiliation{ 
    \institution{European Space Agency (ESA)}
    \city{Cologne}
    \country{Germany}} 
\email{hello@pauldemedeiros.nl}

\author{Nicolas Daniel Herzberger} 
\affiliation{ 
    \institution{IAW RWTH Aachen}
    \city{Aachen}
    \country{Germany}} 
\email{n.herzberger@iaw.rwth-aachen.de}

\author{Michael Preutenborbeck} 
\affiliation{ 
    \institution{IAW RWTH Aachen}
    \city{Aachen}
    \country{Germany}} 
\email{m.preutenborbeck@iaw.rwth-aachen.de}

\author{Andreas Gerndt} 
\affiliation{ 
    \institution{ German Aerospace Center (DLR)}
    \city{Braunschweig}
    \country{Germany}} 
\email{andreas.gerndt@dlr.de}

\author{Frank Flemisch} 
\affiliation{ 
    \institution{IAW RWTH Aachen}
    \city{Aachen}
    \country{Germany}} 
\email{f.flemisch@iaw.rwth-aachen.de}

\author{Florian Dufresne} 
\affiliation{ 
    \institution{Arts et Métiers Institute of Technology}
    \city{Laval}
    \country{France}} 
\email{florian.dufresne@ensam.eu}

\author{Georgia Albuquerque} 
\affiliation{ 
    \institution{German Aerospace Center (DLR)}
    \city{Braunschweig}
    \country{Germany}} 
\email{georgia.albuquerque@dlr.de}

\author{Aidan Cowley} 
\affiliation{ 
    \institution{European Space Agency (ESA)}
    \city{Cologne}
    \country{Germany}} 
\email{aidan.cowley@esa.int}



\renewcommand{\shortauthors}{Bensch et al.}

\begin{abstract}

Future crewed missions to the Moon will face significant environmental and operational challenges, posing risks to the safety and performance of astronauts navigating its inhospitable surface. Whilst head-up displays (HUDs) have proven effective in providing intuitive navigational support on Earth, the design of novel human-spaceflight solutions typically relies on costly and time-consuming analogue deployments, leaving the potential use of lunar HUDs largely under-explored. This paper explores an alternative approach by simulating navigational HUD concepts in a high-fidelity Virtual Reality (VR) representation of the lunar environment. In evaluating these concepts with astronauts and other aerospace experts (n=25), our mixed methods study demonstrates the efficacy of simulated analogues in facilitating rapid design assessments of early-stage HUD solutions. We illustrate this by elaborating key design challenges and guidelines for future lunar HUDs. In reflecting on the limitations of our approach, we propose directions for future design exploration of human-machine interfaces for the Moon.

\end{abstract}



\maketitle
\section{Introduction}

Five decades after the Apollo astronauts last walked on the Moon, humanity now stands at the dawn of a new era in lunar exploration. Spearheaded by the Artemis program, a new generation of crewed lunar landings is set to commence in the near future targeting the Moon's south pole \cite{Smith2020}. Apart from simply furthering our scientific knowledge, a key goal of Artemis is to establish a sustainable human presence on the Moon, with permanent habitats eventually erected on its surface \cite{weber2021artemis}. This endeavor is envisaged to drive the advancement of new astronaut support technologies, ultimately paving the way for human expansion to Mars and beyond \cite{horneck2003humex}. 

The realization of this vision will hinge on the ability of future lunar ground crews to carry out extravehicular activities (EVAs, commonly known as \textit{spacewalks}) of unprecedented scope and complexity, ranging from scouting and geological observations to construction and maintenance of surface infrastructure \cite{braly_augmented_2019,haney2020apollo,landgraf_lunar_2021}. The astronauts undertaking such tasks will have to navigate the desolate and achromatic lunar landscape, characterized by extreme lighting conditions, while grappling with the constraints of their bulky space suits, limiting their mobility and field of vision \cite{davis2019testing}. This has been foreseen to cause high fatigue, poor situational awareness, and impaired spatial orientation \cite{davis2019testing, belobrajdic2021planetary}. Furthermore, due to communication delays between Earth and the Moon, along with frequent radio blackouts, a high level of astronaut autonomy will likely be required \cite{chappell2017evidence}. 

Confronted with these challenges, a promising avenue of exploration in this domain centers on the use of head-up displays (HUDs) along with relevant technologies, such as Augmented Reality (AR), to superimpose contextually relevant information over the user's view of the real world \cite{coan_exploration_2020}. These solutions have been suggested to offer benefits across various aspects of human spaceflight, including astronaut navigation \cite{de_medeiros_categorisation_2022, cardenas2021aaron}.  
HUD solutions have already been successfully utilized to support human navigation in challenging environments on Earth, such as by enhancing the perception and situational awareness of soldiers \cite{livingston_military_2011, mao2019augmented} or by supporting the orientation of drone pilots and vehicle drivers \cite{kaul2017haptichead, bielecki2020enhancing}.
Given these benefits, HUD technologies are also planned to be integrated into future spacesuit models \cite{ILCDover2023PlanetarySuits, scientificamericanAstronautsWill}. Nevertheless, the potential use of HUDs in the context of lunar surface navigation remains a poorly understood domain.  

This limited empirical investigation is largely due to the challenges inherent in replicating a lunar environment conducive to experimental testing. While various emerging lunar technologies have been evaluated in \textit{analogue} test environments, such as aquatic scenarios to mimic reduced gravity \cite{Dorota, Imhof2017ProjectM}, these methods are often associated with significant time-investments, logistical complexities, and high costs \cite{gruber2020amadee}. 

To circumvent these limitations, we have formulated a novel methodology centered on digital twinning that allows for the precise emulation of relevant lunar conditions within an analogue environment simulated using Virtual Reality (VR). This provides HCI practitioners with a cost-effective tool to revisit Apollo-era design challenges, opening them up to modern rapid prototyping and participatory design procedures. 

Following this approach, we have conducted a series of expert-driven evaluations to assess conceptual HUD interface configurations for lunar surface navigation. This paper details our findings from 25 highly qualified participants, including experienced astronauts, instructors, lunar scientists and aerospace engineers.

In conducting this work, we offer a threefold contribution: 
\begin{itemize}
\item We assess the general efficacy of employing VR-based simulated analogue environments for the purpose of evaluating early-stage design solutions in support of human operations on the Moon. 
Drawing on a literature review of historical and contemporary HUD design work, our quantitative and qualitative findings demonstrate the capacity of simulated analogues to surface trends consistent with those observed in real-world experimental deployments. 

\item We present an interview study with uniquely experienced space experts who share their reflections on key design challenges and requirements surrounding future HUD solutions for lunar surface navigation. Their insights reveal a rich problem landscape, laying the empirical groundwork to inspire future design efforts within the emerging \textit{space HCI} domain. \footnote{Several recent initiatives have explicitly called for greater involvement of HCI tools and methods in the context of human spaceflight, see e.g., \cite{pataranutaporn2021spacechi, pataranutaporn2022spacechi}.} To illustrate this, we propose a set of design guidelines for HUD interfaces tailored to the lunar context.

\item In considering the limitations of our approach, we outline a set of recommendations for future VR environments designed to facilitate simulated analogue field studies. These include emphasizing the user's sense of embodiment, incorporating passive haptics, and recognizing the overall utility of multimodal simulation.

\end{itemize}

The remainder of this paper is organized as follows: Section 2 presents challenges faced by astronauts navigating on the Moon and relevant solutions that have been used historically. Moreover, we reflect on navigational interfaces that have been explored by relevant HCI research. In section 3 we detail the methodology of our study, followed by an analysis of the findings in section 4. We discuss the significance of our findings in section 5 and propose directions for future research to expand on our study and further explore HUDs in support of human exploration of the Moon.

\section{Related Work}

\subsection{Human Navigation on the Moon}

As the first human expeditions to the Moon, the Apollo missions pursued a diverse set of scientific and exploratory objectives. Relevant tasks often required the astronauts to embark on lengthy traverses across the inhospitable lunar landscape \cite{eppler_lighting_1991, miller2017operational}. Such EVA operations brought forth considerable challenges, adversely affecting both astronaut safety and performance. 

The most immediate obstacle hampering lunar navigation stemmed from the distinct lighting situation on the Moon. Apollo astronauts reported difficulties arising from poor color contrasts, intense surface reflections, and the absence of shadows in sunlit areas. Conversely, areas covered in pitch-black sharp shadows (a result of the Moon's lack of atmosphere) complicated object identification and obstacle avoidance, making tripping hazards an ever-present concern \cite{eppler_lighting_1991}. The absence of atmospheric light scattering likewise contributed to impaired distance and size estimation, making objects frequently appear closer than they actually were \cite{apollo11_1969, fassett_effective_2020, kagey2022eva, aldrin1969apollo}. 

NASA’s Preliminary Science Report for the Apollo 11 mission described the lunar terrain as featuring "steep slopes, deep holes, and ridges." The lack of natural vertical features, combined with unclear horizon definition and weak gravity cues at the observer's feet, were also blamed for causing "difficulties in identification of level areas when looking down at the surface" \cite{apollo11_1969}.

These issues were further complicated by the Apollo spacesuits. By shifting the astronaut's center of mass higher and further back, the suits made it harder to discern slopes on the surface \cite{aldrin1969apollo}. Similarly, the spherical helmet design limited and distorted the astronaut’s field of vision, thus compounding existing spatial orientation issues \cite{mchenry2020design}.

The lower gravity levels resulted in a situation where astronaut’s suited mobility on the Moon was seen as comparable to unsuited mobility on Earth \cite{apollo11_1969}. Nevertheless, the prolonged duration of many EVA operations resulted in frequent instances of physical exhaustion with astronauts exhibiting increased heart and metabolic rates, necessitating unplanned rests and deviations from planned routes, ultimately compromising mission timelines \cite{marquez2006mission}. 

In summary, extended lunar EVAs thus became synonymous with elevated levels of both mental and physical fatigue, contributing to reduced decision-making abilities and situational awareness, ultimately further compromising the crew’s navigational ability \cite{mullin1960some, petit2019local, endsley1995out,endsley2017direct}.

To overcome these challenges, in addition to preflight information about the lunar surface \footnote{Pre-existing information utilized in support of lunar navigation included Earth-based visual observations, lunar surface photographs obtained by the U.S. Ranger and Lunar Orbiter spacecraft, data from the Soviet Luna 9 and Luna 13 spacecraft, as well as findings derived from the U.S. Surveyor spacecraft }\cite{apollo11_1969}, Apollo astronauts relied on a wide range of tools to support their navigation. On-board the lunar landing module, they determined their spatial position by visually observing stars or constellation landmarks using the \textit{Alignment Optical Telescope}, whilst employing the \textit{Space Sextant} to take navigation readings \cite{mindell2011digital}. Similarly, the \textit{Sun Shadow Device} was used to infer the crew’s location on the lunar surface by measuring the length and direction of the landing module's shadow \cite{he2011integration}. This information was instrumental in planning the astronauts' activities on the lunar surface.

\begin{figure*}[htp]
    \centering
    \captionsetup{justification=centering}
    \includegraphics[width=\linewidth]{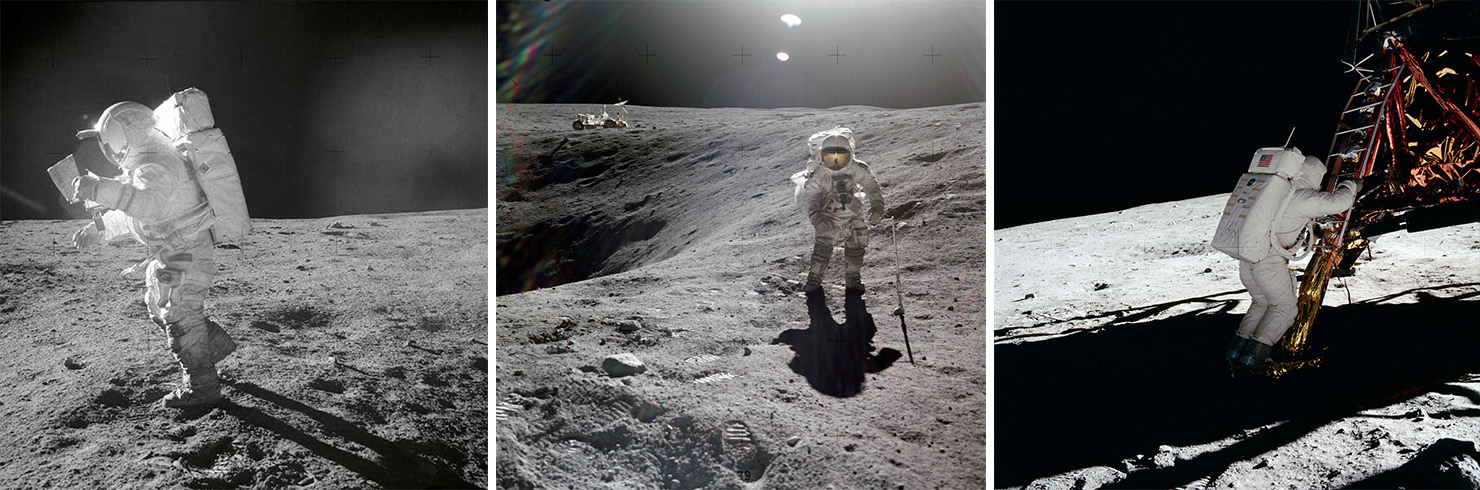}
    \caption{Astronaut Edgar Mitchell using a paper traverse map to navigate on the lunar surface during Apollo 14 (left). Astronaut Charles Duke on the edge of the Plum Crater (middle). Buzz Aldrin egressing the lunar lander during Apollo 11 (right). The pictures underscore the difficult terrain and lighting conditions that astronauts need to face during EVAs on the lunar surface, such as pitch-black shadows and blinding sunlight. Credit: NASA.}
    \label{fig:Apollo}
\end{figure*}

During EVAs, astronauts made use of wrist-mounted watches to time their actions. The watch was likewise used as a tachymeter to measure speed based on the time taken to travel over a known distance \cite{newson2015fifty}. Furthermore, the \textit{Sun Compass} allowed the astronauts to establish a basic sense of direction based on the Sun's position \cite{blucker1971method}. Procedural manuals, detailed checklists, and traverse paper maps likewise saw extensive use \cite{hersch2009checklist, marshburn2003independent}. Far from ideal, the maps, which only offered a top-down perspective, necessitated complex mental rotations and landmark referencing, adding additional cognitive load to the already mentally demanding conditions \cite{anandapadmanaban_holo-sextant_2018} (see Figure \ref{fig:Apollo}).

Despite the vital support provided by these tools, navigating the lunar surface remained a complex and risky endeavour, with success far from guaranteed. Notably, the Apollo 14 mission saw astronauts Alan Shepard and Edgar Mitchell embarking on a geological sampling mission near the Cone Crater. Following a two-and-a-half-hour hike up the steep hill to the crater’s rim, the astronauts had to abort their mission due to not being able to locate their destination, citing “non-optimal sun exposure and limited surface visibility” as contributing factors \cite{carr2003geologic}.


Due to the presence of strategically important resources, such as water ice conserved in perpetually shadowed craters, the upcoming Artemis Moon missions will center on the lunar South Pole \cite{Smith2020}. Characterized by low sun angles that leave much of the landscape shrouded in darkness, this geographical focus, coupled with the ambitious objective of establishing and sustaining permanent human outposts, will inevitably further amplify the challenging nature of lunar EVAs. 

Against this backdrop, it is clear that a robust navigation toolset will be essential for future lunar astronauts. In the next section, we will thus discuss how contemporary HUD interfaces may have the potential to help address many of these challenges effectively.


\subsection{Navigation HUDs} 

\label{interfaces}

Navigation interfaces and related technologies have seen substantial evolution since the Apollo era. While a comprehensive review would be beyond the scope of this paper, below we focus on key interface paradigms that have shown particular prevalence in the context of navigational support. This focus is well aligned with existing research, such as that of Syberfeldt et al., who classified support displays into a \textit{hand-worn}, \textit{head-worn} and \textit{spatial} category \cite{syberfeldt2016support}.

\subsubsection{Head-down and suit-mounted displays}
The aviation industry has been at the forefront of pioneering new navigational interfaces, introducing, for instance, the \textit{Multi-function Displays} (MFDs) capable of displaying diverse navigational data, including real-time terrain topography maps, surrounding air traffic information, waypoints, planned routes, and course deviation \cite{ogan2016navigational}. Similarly, the automotive domain has helped popularize head-down displays integrated into a vehicle’s dashboard, featuring maps and other navigational tools \cite{yeh2003head}. Such displays have been shown to consistently improve navigational precision for pilots and drivers \cite{ogan2016navigational}. 

In contrast, contemporary astronauts embarking on EVAs in Low Earth Orbit while wearing the \textit{Extravehicular Mobility Unit} (EMU) spacesuit only have access to limited real-time information provided via simple text displays and verbal support (by mission or ground control) \cite{hodgson2003requirements}. A paper checklist, approximately 50 pages long, is used to access information about the EMU and its suit-mounted \textit{Display and Control Module} (DCM). The DCM requires astronauts to look down at an awkward angle, often necessitating a Fresnel lens for better visibility due to the short viewing distance, and a wrist mirror to read reflected images of controls located on the DCM's front. It provides limited information (12 characters) and frequently intrudes into the astronaut's primary workspace, further complicating the access to information during EVAs \cite{gernux1989helmet}.

In response, NASA has begun investigating the use of electronic displays that could be attached to the wrist or the front of the astronaut suit. This led to the development of an electronic checklist designed to enhance productivity by providing easy access to a comprehensive database \cite{gernux1989helmet}. However, the prototype electronic cuff checklist used in four Shuttle flights faced several issues, including glare, low contrast, and interference with the work envelope \cite{marmolejo1996electronic}. In comparable initiatives, such as Project Moonwalk, a chest-mounted display featuring a touchscreen, alongside a wrist display intended to replace the traditional cuff checklist, were implemented. These devices were utilized to present mission procedures during two analog EVAs, conducted in environments replicating Mars and underwater scenarios \cite{imhof2017project}.

However, such digital displays have likewise faced criticism for necessitating users to transfer their attention back and forth between instructions and the locus of their action, which may lead to a higher mental workload and a decrease in situational awareness \cite{tang2003comparative}.

\subsubsection{Wearable HUDs}
Due to their ability to seamlessly integrate real-world views with computer-generated data directly in the user's field of vision, the use of HUDs has emerged as a frequently discussed alternative \cite{milgram_taxonomy_1999}. For example, \citet{bark2014personal} demonstrated that car-based HUD systems featuring navigation arrows projected on a see-through display in front of the driver help maintain the driver’s focus on the road, resulting in improved performance and safety \cite{schneider2021navigation}. Similar findings have been reproduced with a 2D minimap projected into the corner of the driver’s view \cite{ma2021does}.

In the context of human spaceflight, various tests investigated the use of wearable computers with near-eye display prototypes, such as the WearSAT, to assist in task execution \cite{carr2002wearable}. The Haughton-Mars Project (HMP) explored the use of two different HUD prototypes in exploration and field science \cite{boucher2002investigation}.

In the mid-to-late 1980s, NASA researchers explored similar uses of Head-mounted displays (HMDs), eventually developing experimental see-through head-mounted screens displaying text, graphics, and video \cite{gernux1989helmet}. The Hamilton Standard display, for example, was designed to provide information about checklists, navigational aids, shuttle transmission video, and warnings about the EMU status. However, the high costs of integrating HMDs into the EMU led to their deferral as a consideration for future generations of spacesuits \cite{gernux1989helmet, marmolejo1994helmet}.

\subsubsection{Spatial HUDs}
In an evolution of traditional HUDs, head-worn displays featuring tracking capabilities have been employed to spatially overlay points of interest in the real world with AR navigational cues. Research has demonstrated a number of benefits to such a practice, including enhanced performance (likely due to a reduced visual scanning time and reduced visual clutter, enhancing the integration of the information in the mental model of the user) and reduced attention capture \cite{wickens1995object,wickens2003aviation,mccann1996scene}.

Navigational information integrated into the environment as a form of augmented reality information has also been found to significantly reduce mental workload and driving error, particularly in ambiguous driving situations \cite{bauerfeind2019does}. Moreover, the AR interface was perceived as being more comprehensible and was overall preferred by the participants, in contrast to information displayed as a HUD overlay without being superimposed on the real environment. Similarly, research suggests that AR-based navigation cues reduce the cognitive effort to map navigational information with the real world, thus enhancing driving performance \cite{kim2009simulated}. Spatial elements have likewise been shown to improve the overall user experience of drivers by, for instance, projecting semi-transparent guiding pathways onto the road \cite{schneider2021navigation,narzt2006augmented}. By projecting 3D grid meshes on top of real environments, similar solutions were also found to improve the user’s understanding of distances and terrain structure \cite{lopez-contreras_testing_2022,mao2019augmented} as well as enhancing drivers’ vision in adverse weather conditions \cite{charissis2009interface}. Similarly, virtual “tunnels in the sky” guiding pilots to their destination were found to improve their situational awareness and reducing mental workload (e.g., \cite{prinzel2004efficacy, martins2020evaluation}). 

As of yet, no such spatial navigation solutions have however been developed for astronaut EVAs. Nevertheless, some prototypes have found their way into analogue testing and training activities. For instance, the Microsoft HoloLens 2 HMD has been employed to provide navigational guidance during an analogue mission near the Kilauea volcano \cite{anandapadmanaban_holo-sextant_2018}. Notably, participants in the study voiced their preference for the experimental HUD over traditional navigational devices, citing its superior legibility under direct sunlight. 

In the same analog mission setting, the feasibility of utilizing HUDs to visualize optimal traversal paths in challenging terrains was demonstrated by \citet{anandapadmanaban_holo-sextant_2018}. Building on this foundation, Bonilla et al. proposed a prototypical AR-based obstacle avoidance system \cite{bonilla2022moonbuddy}.

\subsubsection{Global Orientation HUDs}
While the aforementioned interfaces are primarily concerned with supporting the user’s understanding of the immediate surroundings, providing them with precise instructions ensuring that they effectively and safely maneuver through their current space, another emerging use of HUDs focuses on providing users with a ‘global’, strategic awareness of their operational context, such as by visualizing a comprehensive mission plan or geographical information \cite{tonnis2005experimental, barfield1995situation}. 

For instance, military HUDs, incorporating compass tools that project cardinal directions, heading, and target location into a soldier's field of view, have been shown to significantly reduce mission completion times while simultaneously increasing situational awareness by visualizing waypoints and the position of allied forces \cite{bielecki2020enhancing, colbert2005augmented}.

Similarly, studies of airplane pilots have demonstrated that relying on such global information may result in greater route deviation but generally contributes to a decrease in workload by shifting the pilot’s attention to the bigger picture, rather than every small deviation \cite{beringer1999flight}. Such global orientation interfaces have likewise been found to decrease the risk for cognitive tunneling while increasing the pilot’s attention to the outside world \cite{mccann1993attentional,wilson2002comparing}. 

\subsection{Designing for the Moon}

In summary, advancements in HUD and AR technologies are showing promise in aiding human navigation during complex operations, with features such as suit-mounted displays, minimap widgets, various spatial AR visualizations, and compass tools having already shown promising results across several fields. Wearable HUDs have consistently demonstrated, for instance, a reduction in perceived task load and improved usability compared to traditional suit-mounted and handheld displays. Such trends appear to be even more prominent with spatial HUDs. 

Against this backdrop, several initiatives in the human spaceflight domain have set out to explore potential navigational interfaces in support of future astronaut EVAs on the Moon. The need for representative testbeds to facilitate such exploration has, however, resulted in most studies taking place in real-world analogue sites, such as neutral buoyancy facilities (large "swimming pools" that mimic low or zero gravity conditions underwater), or natural environments, such as deserts or cave systems that emulate relevant lunar conditions (e.g. soil and lighting scenarios) \cite{bessone2015training}. 

In a recent study, for instance, \citet{ahner2023developing} developed an AR HUD incorporating a minimap, and spatial color-gradient pathway terrain overlay and a compass. This interface was subsequently assessed via experimental deployments at NASA’s Johnson Space Center (JSC) Rockyard. Although comprehensive evaluation metrics were somewhat limited, preliminary user feedback described the features as "intuitive and straightforward". The color-gradient path, in particular, was praised for aiding users in distance estimation, although no comparative analysis was provided between the various navigational aids.

Similarly, the Pathfinder application guided astronauts through intricate virtual terrains using a minimap and AR pathway-based visual markers \cite{anastas2020augmented}. The system also featured a compass widget displaying cardinal directions and waypoint indicators. Initial findings, albeit based on a small sample size (N=6), indicated varying impacts on user performance and cognitive workload when comparing guiding AR pathways and waypoint visualizations. The AR pathway resulted in superior path-finding performance, while waypoint-based visualizations were associated with reduced mental workload. No data was however presented regarding the usability of the compass and minimap features.

The reliance on analogue testbeds in such studies has attracted criticism for their logistical complexity and resource-intensive nature, making experimental deployments prohibitively costly for many organizations \cite{nilsson2023out}. Consequently, research in this area has been limited, making the design of navigational HUDs for lunar operations a poorly understood domain. 

In the following section, we outline our approach to addressing this deficit by engaging a diverse group of domain experts and immersing them into a VR-based simulated analogue environment featuring hypothetical HUD elements. With that we seek to open up lunar HUD design to established HCI practices, ultimately helping to develop a more holistic perspective on the challenges, opportunities, and design requirements (and resulting guidelines) surrounding HUD technologies in service of human navigation on the Moon.

\section{Methodology}

Faced with the challenge of facilitating user operations in extreme environments, the United States military introduced the Human-Systems Integration (HSI) engineering approach in the late 1970s \cite{drillings2015human}. Subsequently adopted and popularised by NASA for the purposes of space systems development, HSI is characterized by combining elements of human factors and HCI, placing focus on a holistic consideration of human capabilities and limitations in designing, implementing, and operating hardware and software \cite{boy2021human, booher2003handbook,flemisch2021towards}. Such practice is typically guided by concurrent assessments of mission-critical elements in an analogue environment providing a representative operational context \cite{howe2013nasa}.

In an attempt to bring the analytical lens of HSI into the ideation and early-stage prototyping activities, Flemisch et al. have recently proposed a \textit{Human Systems Exploration} paradigm as an iterative design approach for innovative human-technology systems \cite{flemisch2022human}. By progressively structuring the design-, use-, and evaluation space, Human Systems Exploration seeks to bridge the gap between more creative design methodologies, such as Design Thinking, and more conventional development methods from systems and human factors engineering \cite{meyer2021increasing}. The approach advocates a holistic consideration of different development methods (e.g. qualitative/quantitative, slow sectional/cross-sectional, subjective/objective) and is also commonly described as a \textit{balanced analysis} \cite{flemisch2019making}. 

Our methodological orientation adopts and expands on the Human Systems Exploration paradigm by examining the potential viability of a VR-based simulated analogue testbed as a means to concurrently and collaboratively explore the interplay between relevant constraints on human navigation in a lunar environment (such as poor lighting conditions or absence of landmarks) and conceptual HUD solutions.

\subsection{Virtual Testbed}
To provide a realistic context for our HUD exploration, we employed VR technology to simulate a realistic lunar analogue environment. All lunar conditions and elements of mission architecture featured in this analogue were guided by the Concept of Operation (ConOps) \cite{beaton2019using} for initial phases of humanity’s return to Moon outlined by  Landgraf et al. \cite{landgraf2021lunar} in collaboration with several international space agencies. 

\begin{figure}[htp]

\includegraphics[width=1\linewidth]{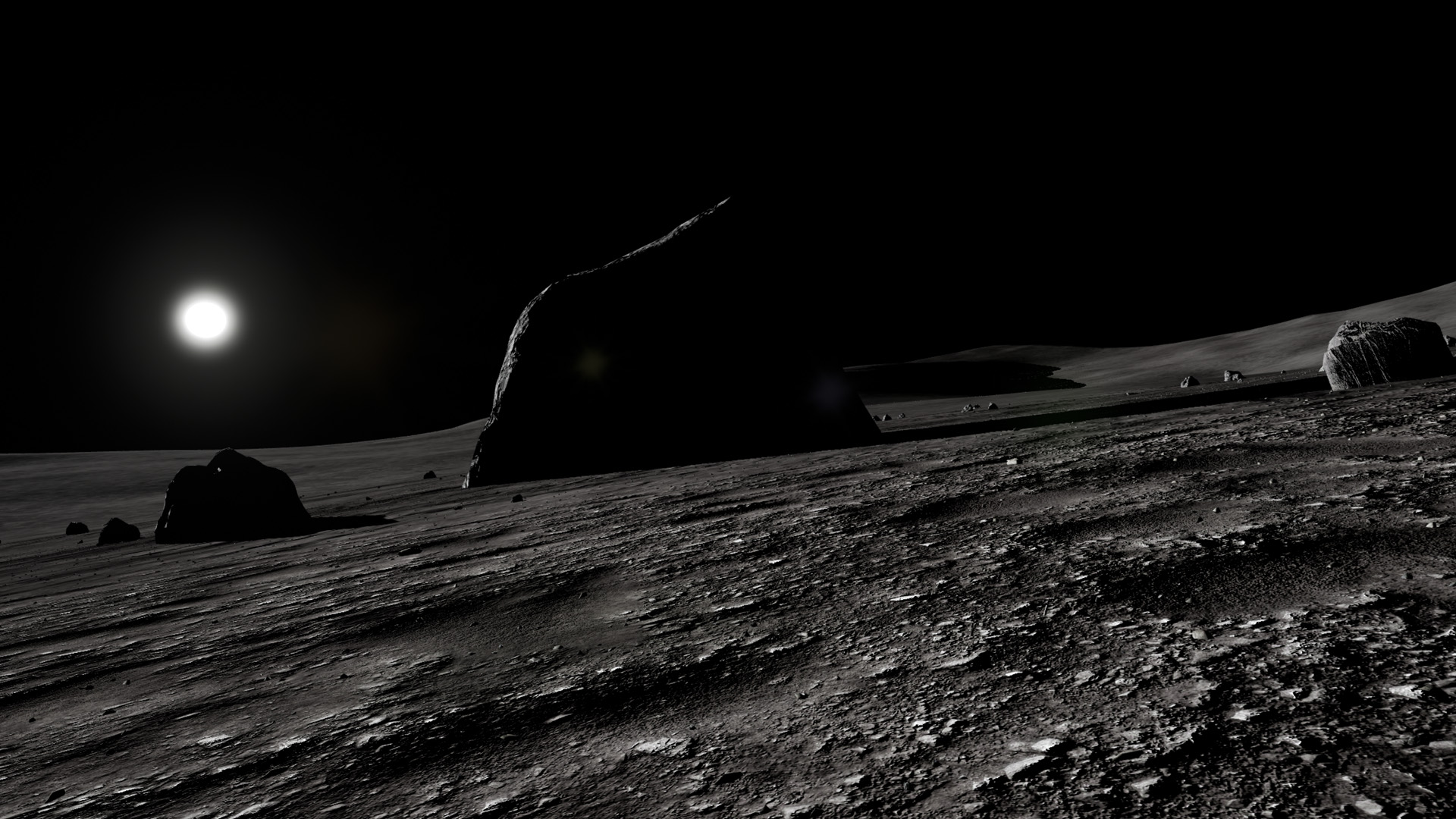}
\caption{Screenshot of the virtual environment that was utilized during the study, displaying accurate lighting reflections and shadows.}
\Description{Figure 3 showcases a rendering of the virtual lunar environment employed in our study, as viewed from an astronaut's first-person perspective. It shows elevated terrain, with the Sun shining in the background, while some parts of the terrain are located in pitch-black shadow.}

\label{fig:VirtualMoonscape}

\end{figure}

Leveraging topographical data obtained from the Lunar Reconnaissance Orbiter \cite{smith2017summary}, we digitally reconstructed \(64 \, \text{km}^2\)
of the Shackleton Crater region at the Moon's south pole \((89.9^\circ \text{S}, 0.0^\circ \text{E})\) - a probable destination for the coming Artemis human landings \cite{smith_artemis_2020}. Given the 100-meter pixel resolution of the original topographic scans, finer elements such as boulders had to be incorporated manually using the Cinema 4D modeling software. The simulated lunar terrain was then procedurally textured using bespoke terrain shaders (see Figure \ref{fig:VirtualMoonscape}). 

To simulate elements of potential crewed mission procedures, several prospective landing systems were also positioned in the environment, including the Argonaut cargo delivery lander \cite{landgraf2022autonomous} and the Starship HLS \cite{seedhouse2022starship}. All 3D models utilized in the study were textured with physically-based rendering materials to accurately replicate light interactions, such as glare and reflections.

To enhance the perceived authenticity of the virtual testbed, participants were virtually embodied in an xEMU EVA suit \cite{cardenas2021aaron}, complete with a headlamp-equipped helmet. The suit's torso and gloves were rendered visible to the user, who also cast a full-body shadow. Particle effects were also employed to visually simulate lunar dust plumes upon physical interaction with the lunar surface.

The sun was placed in the direction of north and its intensity was set to $1.37 kW/m^2$ to mimic the conditions on the lunar south pole \cite{vanoutryve2010analysis}. All forms of indirect lighting and light scattering were disabled to recreate the pitch-black shadows stemming from the lack of lunar atmosphere.

The virtual scene was instantiated utilizing the Unreal Engine 4.26 game engine. Participants were interactively immersed in the virtual environment using a HTC VIVE Pro 2 HMD rendering the virtual environment at a refresh rate of 90 Hz with a resolution of 2880 × 1600 pixels (1440 × 1600 pixels per eye) and providing the user with a diagonal field of view of 110 degrees. Two HTC Vive controllers were used for navigation, with only continuous movement (rather than teleportation) being implemented to avoid potential interference with the navigation task. The VR experience was deployed on a desktop computer with an RTX 4090 graphics card, ensuring a consistently smooth framerate.

\subsection{Interface Configurations}

In order to stimulate discussion and elicit relevant reflections, we utilized our simulated analogue environment to reenact the functionality of several hypothetical navigational elements. Guided by our literature review of existing paradigms in navigational HUD interfaces (see section \ref{interfaces}), 
we included four distinct interface configurations: 

\begin{itemize}
    \item \textbf{Suit-mounted tablet} was based on concepts such as the Electronic Field Book \cite{turchi2021system}, developed by the European Space Agency. Its wrist-mounted nature of the \textit{tablet} meant that it offered a usability experience that was comparatively closer to traditional maps used by the Apollo astronauts, which, in turn, provided us with the means to compare HUD solutions to more conventional navigational tools. The map offered a bird's eye perspective of the EVA area relevant to the navigational task. A blue line indicated a recommended traverse route to reach navigation targets. An arrow symbol was used to denote the participant's current position and orientation. The tablet was secured to the participant's arm, making it necessary for them to raise their arm and align the tablet with their line of sight to use it. 

    \item \textbf{Wearable HUD}, displayed at the center bottom of the user’s view, provided a top-down \textit{minimap} representation of the surrounding environment via a live camera feed. A blue line indicated the recommended traverse route. The participant's location was symbolized by an arrow, and the map's orientation was adjusted dynamically in real time to mirror the participant's movements and changes in direction. 

    \item \textbf{Spatial HUD}, similar to the one utilized by Anandapadmanaban et al. \cite{anandapadmanaban_holo-sextant_2018} and  Anastas et al. \cite{anastas2020augmented} shows the recommended traverse route guiding the user from the starting position to their destination. The \textit{AR Pathway} was visually depicted as a light, semi-transparent blue pathway projected on the ground. The pathway possessed a degree of thickness, ensuring it remained visible when viewed from a lateral perspective. 

    \item \textbf{Global orientation HUD} displaying cardinal directions and featured a blue flag, which indicated both the direction and distance (in meters) to the next waypoint along a recommended traverse route based on the user's location in the environment. To ensure correct orientation of the \textit{compass}, the waypoints were placed one meter apart along the recommended route. If the flag moved out of the user’s field of view, an arrow would appear in the HUD, indicating the direction of its position. 

\end{itemize}

In addition, each interface solution had the elapsed EVA time, the distance to traverse destination, as well as the traverse progress in percentage displayed in the user’s field of view. The four conditions were investigated following a within-subjects design.

Following the Human Systems Exploration approach, our aim was not to assess the engineering viability of any particular interface design but rather to explore key human requirements and preferences pertinent to lunar navigation. As such, we did not take into account any technological limitations, e.g., the power consumption or geolocation constraints inherent in contemporary technology. No lunar positioning system, for instance, is in operation as of yet (the prospective implementation of such systems is, however, a frequent topic of debate \cite{wijnen2018cubesat, akimov2020navigation, giordano2022lunar}). 

Neither of our navigational solutions can thus be seen as technically correct, and nor are they meant to be. Rather, in a manner analogous to critical design \cite{bardzell2013critical}, they were intentionally designed to engender major paradigms in navigation interface technology and thus help surface essential user reflections and considerations that are bound to impact the adoption of future lunar navigational solutions, regardless of their eventual design or implementation. 

\begin{figure}[htp]

\includegraphics[width=1\linewidth]{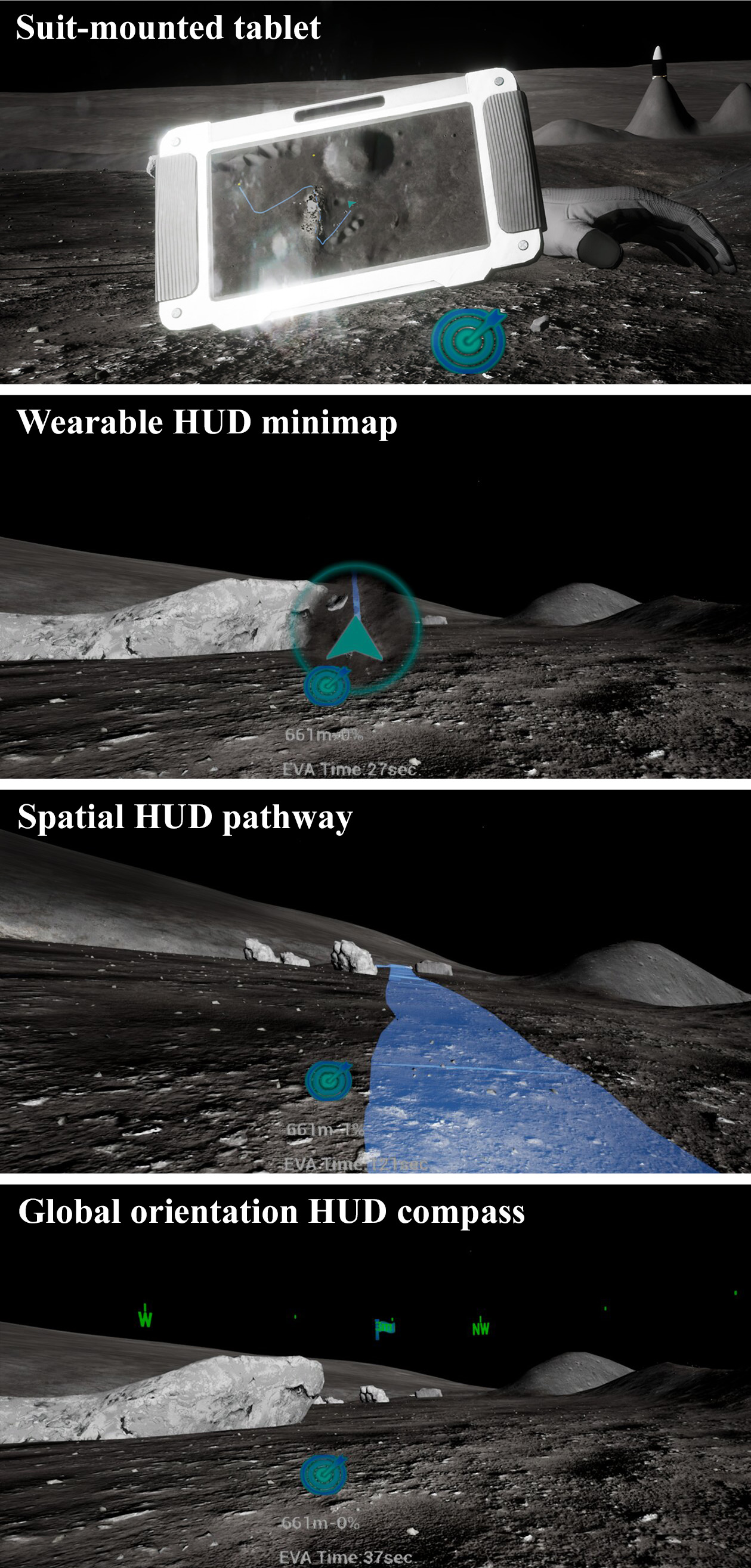}
  \caption{Screenshots of the distinct navigational interface conditions featured in the study.}
  \label{fig:Navigation}
  \Description{Figure 2 provides a comparative visual overview of the four experimental conditions (Tablet, Minimap, AR Pathway, and Compass). The depicted terrain is characterized by its relatively flat topography, punctuated by rocks, and is illuminated.}

\end{figure}

\subsection{Participants}
Given the specialized nature of our topic of enquiry, focusing on future Artemis missions that are already planned to begin during the next years, we hand-picked and extended personal invitations to relevant domain experts, seeking to include the perspectives from a range of domains related to human spaceflight and lunar science. 

Overall 25 domain experts (Age: \textit{M}= 38.8 years, \textit{SD}= 9.8 years) were recruited for the study (see table \ref{tab:participants}). Three astronauts were included. Astronauts 1 and 2 are both active and have together logged around 547 days in space, including 14 hours of EVA operations outside the International Space Station (ISS). Astronaut 3 is a recent recruit currently undergoing basic training. The engineers have all experience supporting space missions and training activities. The instructors have all experience with delivering astronaut training, some of them also being involved in overseeing astronaut operations through EUROCOM - the operators responsible for communication with the ISS. Notably, Instructor 5 is an expert in astronaut EVA training, with experience designing and evaluating different types of spacesuits. Most of the scientists are involved in field study activities for future lunar and Martian exploration missions, while others are working on topics related to Moon exploration. Finally, the subject matter experts were quite experienced using VR technologies as demonstrated by the self-reported scores, \textit{M}=3.27 (\textit{SD}=1.35), assessed on a scale from 1 (no experience) to 5 (highly experienced).

\begin{table*}[]
\small
\centering
\begin{tabular}{|l|l|p{0.35\textwidth}|p{0.42\textwidth}|}
\hline
\textbf{Role} & \textbf{Gender} & \textbf{Job Title} & \textbf{Area Of Expertise} \\
\hline
Astronaut 1  & F & Astronaut & Human Spaceflight, EVA \\
Astronaut 2  & M & Astronaut & Human Spaceflight, EVA \\
Astronaut 3  & M & Astronaut Candidate & Human Spaceflight, EVA \\
Engineer 1 & M & Aerospace Engineer & Aerospace Engineering \\
Engineer 2  & M & Engineer/Roboticist & Engineering, Robotics, XR \\
Engineer 3 & M & Computer Systems Engineer & Computer Science and Systems Engineering \\
Engineer 4 & M & Aerospace Engineer & Aerospace Engineering, Astronaut Training \\
Engineer 5 & M & Engineer &  Human Factors and Ergonomics \\
Engineer 6 & M & Aerospace Engineer & Aerospace Engineering \\
Engineer 7 & M & Procedure Manager/Operations Engineer & Procedure Management and Operations Engineering \\
Engineer 8 & F & Instructor/Eurocom/ECOS Ops Engineer & Instruction, Eurocom, ECOS Ops Engineering \\
Engineer 9  & M & VR/AR Software Engineer & VR/AR Software Engineering \\
Engineer 10 & M & In-Space Manufacturing Specialist & In-Space Manufacturing \\
Instructor 1 & M & Astronaut Instructor/ECOS-OPS & Astronaut Instruction, ECOS-OPS \\
Instructor 2 & M & Astronaut Training and Simulations Support & Support to Astronaut Training \& Simulations of EVAs \\
Instructor 3 & F & Training and EUROCOM Specialist & Training and EUROCOM \\
Instructor 4 & F & Spaceflight Operations Specialist & Spaceflight Operations \\
Instructor 5 & M & EVA Operations Trainer & EVA Operation Training, Lunar EVA Simulation \\
Scientist 1 & M & Computer Scientist & Science and Computer Science \\
Scientist 2 & F & Research Fellow & Machine Learning \\
Scientist 3 & M & Lunar Scientist & Lunar Science, LUNA Facility, Aerospace Engineering \\
Scientist 4 & M & AI Researcher & Research in AI \\
Scientist 5 & M & Physicist & Physics, EUROCOM Crew Support \\
Scientist 6 & M & Space Radiation Shielding Specialist & Space Radiation Shielding \\
Scientist 7 & M & Scientific Researcher & Scientific Research and Development \\
\hline
\end{tabular}
\caption{Summary of participant demographics. Indicated areas of expertise were self-reported.}
\Description{Table 1 presents a detailed summary of participant demographics and areas of expertise, organized alphabetically by role. The table is structured into four columns: 'Role,' 'Gender,' 'Job Title,' and 'Area of Expertise.' The 'Role' column categorizes participants into five distinct roles: Astronaut, Engineer, Instructor, and Scientist. The 'Gender' column indicates the gender of each participant, either 'M' for male or 'F' for female. The 'Job Title' column provides the professional title of each participant, ranging from 'Astronaut' to specialized engineering and scientific roles. Finally, the 'Area of Expertise' column delineates the specific domains in which each participant has expertise, such as 'Human Spaceflight, EVA,' 'Aerospace Engineering,' and 'Machine Learning,' among others. }
\label{tab:participants}
\end{table*}

\subsection{Procedure}

Each participant was individually invited to our lab. Upon arrival, they were briefed concerning the purpose and nature of the study and asked for consent to have their session recorded on audio. They were then asked to fill in a pre-task questionnaire, which included questions about their demographic information and their level of prior experience with VR. The session subsequently revolved around two tasks: 

\begin{itemize}

\item \textbf{Task 1:} As a warm-up activity, participants were provided with a paper sketch of an astronaut helmet. They were then asked to draw their envisioned navigation HUD interface for lunar EVAs (see Figure \ref{fig:figure3}). This approach, commonly employed in HCI for rapid prototyping and brainstorming of novel technologies \cite{lewis2019sketching}, provided us with the means to gauge the participant’s general preferences regarding HUD features and their layout. 
Throughout this task, we employed a think-aloud protocol \cite{jaaskelainen2010think} in conjunction with a semi-structured interview to gain insight into the experts' reasoning. 
Apart from serving as an introduction to our study, this task provided participants with an opportunity to articulate their general preferences before encountering our interface configurations. This ensured that our discussions and findings were grounded in participant's own views unbiased by our own preconceived design ideas.
\\
\item \textbf{Task 2:} In the study's second phase, participants were introduced to our virtual testbed with the objective of completing a 500m traverse route from one Argonaut lander to another. They performed this task four times, each time using a different navigation feature. We continued to use the think-aloud protocol to capture their thoughts. To minimize any potential order-related biases, we counterbalanced the interface conditions. At the beginning of each condition, participants were briefed on the condition, along with an opportunity to become acquainted with it. The initial training phase, encompassing the first two percent of the traverse route participants were required to navigate, was later omitted from data analysis.

Post-task, participants were asked to complete a questionnaire assessing the perceived workload and usability of the different navigation features. Specifically, we implemented standardized instruments to quantitatively measure the perceived workload induced by using the navigation feature through the NASA TLX \cite{hart1988development}, and features' usability utilizing the System Usability Scale (SUS) \cite{brooke1996sus}.  A single question derived from a presence questionnaire developed by Slater et al. \cite{slater1993presence} was used to evaluate the psychological feeling of being present in the virtual environment (" On a scale of 1 to 7, please rate your sense of being present in the virtual environment,
where 7 represents your normal experience of being in a place. I had a sense of “being
there” in the lunar environment:").

Finally, as an objective performance measurement, we recorded the deviation, in meters, of the participant from the recommended traverse route and the time that participants needed to complete the route. 
\end{itemize}

\subsection{Data Analysis}

Audio recordings of the study sessions were transcribed and independently coded by two of our researchers. Any inconsistencies were addressed through a discussion and subsequently
synthesized into a qualitative thematic analysis. 

In addition to the qualitative analysis, we also reviewed our participant’s questionnaire responses to derive quantitative findings concerning perceived usability and workload. Following our literature review, we sought to establish whether our simulated analogue study would reveal trends similar to those in comparable real-world studies, namely the introduction of HUDs resulting in a reduction of perceived workflow and improved usability, with spatial interface features further enhancing this trend. While our focus on the criteria of workload and usability is somewhat arbitrary and does not offer a holistic assessment, given the prevalence of these two criteria in existing literature, we might argue their exploration here provides a broadly indicative insight into the general efficacy of simulated analogues. Such insights could prove a useful starting point for future research. In addition, we also enquired about perceived presence, tracked task completion times and route deviation. 
\begin{itemize}

    \item {\textbf{H1:}} Perceived usability of a wrist-mounted tablet during navigational tasks performed in the simulated analogue environment is lower than that of HUD interfaces. The use of spatial HUD interfaces further amplifies this difference.
    
     \item {\textbf{H2:}} Perceived workload associated with the use of a wrist-mounted tablet during navigational tasks performed in the simulated analogue environment is higher than that of HUD interfaces. The use of spatial HUD interfaces further amplifies this difference.
     
\end{itemize}
Repeated Measure Analyses of Variance (rmANOVAS) were conducted for each dependent variable using Greenhouse Geisser correction due to sphericity violations. Post-hoc comparisons were conducted with Bonferroni correction. The significance level $\alpha$ was set as .05.  The statistical analysis was performed using the Python package Statsmodels.

From the total number of 25 participants, 22 were included in the statistical analysis of the quantitative questionnaire data. 3 participants (incl. the two astronauts) had to be excluded, as they did not fill out the questionnaire after the experiment due to time constraints.

17 participants were included in the performance data analysis. Three participants had to be excluded as they did not fully complete the VR traverse task (incl. the astronauts), while 5 participants had to be excluded due to recording errors, resulting in missing data.

The questionnaire and performance data examined in this study remain confidential at this time due to their involvement in an active research project. Following the conclusion of our analysis and the release of additional related research papers, we intend to release the full dataset for public access.

\section{Findings}
 
\subsection{The Need for HUDs}

During the sketching sessions, participants consistently highlighted important strengths and advantages offered by HUD interfaces when compared to more traditional navigation tools. Instructor 4, for instance, commended HUDs for freeing up astronauts' hands for other tasks, while Instructor 1 brought up the convenience of having critical information directly within one’s field of view. Astronaut 1, drawing from her past experience as a military pilot, echoed these sentiments, emphasizing her preference for HUDs over conventional tools:

\begin{quote}
\textbf{Astronaut 1:} "Once you get used to a heads-up display, you never go back. There's no comparison.” 
\end{quote}

The simulated tablet device featured in our study prompted our participants to elaborate their views further. Engineer 5 questioned the ergonomics of using handheld devices while wearing a space suit and helmet visor. Engineer 3 and Astronaut 2 shared a similar perspective, explaining that astronauts using a tablet during EVAs would have to lift up the device whenever they needed to reference navigational instructions, which would be both "cumbersome and fatiguing". Similarly, Astronaut 3 described the tablet as a source of "interruption" due to forcing its users to switch their focus back and forth between the tablet and the environment. Instructor 4 went a step further, calling the tablet a potential safety risk:

\begin{quote}
\textbf{Instructor 4:} "If the tablet is my only navigational aid, then that's a real danger. Because then I will be looking at the tablet all the time and not looking around. That’s not a good idea on Earth. But on the Moon, it can get you killed. So I would definitely not use that as a primary navigational aid.” 
\end{quote}

Despite its evident drawbacks, the concept of handheld navigational tools was not entirely dismissed. Scientist 4 and Instructor 1, for instance, took a more balanced approach, suggesting that handheld devices like tablets could serve as a valuable supplement to HUD interfaces. According to them, HUDs should primarily display crucial real-time data, while handheld devices could be employed to access supplementary, less critical information. 

In contrast, Engineer 3 envisioned the tablet simply as a backup, offering navigational support in case of a HUD malfunction. 

In summary, whilst traditional navigational tools, such as tablets, were not entirely devoid of merits, our participants were in agreement that exploring the potential of navigational HUDs constitutes a legitimate pursuit and that advancing their design will be vital for coming lunar expeditions. The functions such an interface should fulfill and the manner in which these functions ought to be integrated then became the focal point of further discussions.

\subsection{Key Building Blocks of a Lunar HUD}

When our participants were asked about key features of a navigational HUD, the majority stressed the significance of including indicators of vital signs related to both the astronaut and their spacesuit. The importance of indicating oxygen levels, for example, was repeatedly brought up. Other frequently mentioned vital signs included the astronaut's heart rate, internal suit temperature, battery status, connectivity strength, and EVA elapsed time. While not directly tied to navigation, our participants agreed that access to this information would play a pivotal role in guiding their navigational decisions by helping them, for instance, determine whether to adjust their traversal speed, when to take a break, or even assess whether it was necessary to abort an EVA mission.

The idea of incorporating minimaps into the user's HUD was another common suggestion among our participants, often drawing inspiration from conventional GPS or even traditional paper maps. However, upon considering the challenges presented in our virtual lunar landscape, it became apparent that such a map would have to meet several distinctive criteria. 

Criticizing the relatively minimalistic map postulated by our VR simulation, Engineer 3 and Scientist 1, for example, emphasized the importance of maps indicating in real-time which geographical areas are illuminated by the Sun and which are covered in shadow. 

The challenging lighting conditions likewise led most participants to recommend adding elevation and contour data to the map. This, they reasoned, would assist astronauts in identifying steep slopes and other potentially hazardous irregularities in the terrain. Scientist 3 proposed that HUD-based minimaps should be rendered in 3D to enhance the readability of the environment's topology. Engineer 6 went a step further, suggesting integration of LIDAR technology into spacesuits for real-time topographical scans of the astronaut's surroundings.

No matter how well-crafted a map might be, our participants were in agreement that it would need to be complemented by spatial AR elements highlighting relevant terrain features directly in the user’s field of view. Astronaut 1, for example, stressed the vital role of real-time hazard indicators, proposing the use of visual cues (e.g., “colored circles”) to outline potentially hazardous terrain features in the user’s proximity, such as boulders or craters. Similarly, Astronaut 2 pointed out the challenge of accurately estimating distances on the Moon due to the absence of an atmosphere. As a compensatory measure, he proposed the use of AR-based visual distance indicators for relevant points of interest in the astronaut’s field of view. 

Whilst discussing these navigational elements, participants highlighted the importance of adhering to existing color schemes and standards. A frequent suggestion, for example, was the use of red for warning and emergency information. Unsurprisingly, many participants also advised against using white and black colors in HUDs, as these colors dominate the lunar landscape. Astronaut 2 and Instructor 3 stressed the need to maintain consistency with existing solutions but also recognized the importance of innovation that does not compromise usability. Scientist 1, however, cautioned against introducing innovations that could clutter the user's view, advocating for a measured approach: 

\begin{quote}
\textbf{Scientist 2:} "I'd be very careful on the color choices. What you want to avoid is putting out too much information and too many colors around them [the astronauts], distracting them so that they can't see the actual environment around them clearly.” 
\end{quote}

As the sessions progressed, and the number of discussed HUD features grew, similar concerns turned into a recurring theme in our participants’ reasoning. It became evident that creating an effective navigational HUD was not only about incorporating the necessary supportive elements but also about presenting them in an efficient and minimally intrusive manner. This observation then sparked a discussion about the ideal HUD layout.

\subsection{Lunar HUD Layout}

The sketching sessions played a crucial role in helping participants conceptualize and express their preferred HUD layout. 
Astronaut 2, in particular, provided a detailed account of his ideal solution. He divided the helmet view into two distinct sections: the “peripheral vision area” along the edges of the helmet and the “primary focus area” at the center. The peripheral vision area, he argued, ought to be reserved for auxiliary information, such as vital signs indicators, whilst the primary focus area should more directly deal with navigation support. 

\begin{figure*}[htp]

\centering
\includegraphics[width=1\textwidth]{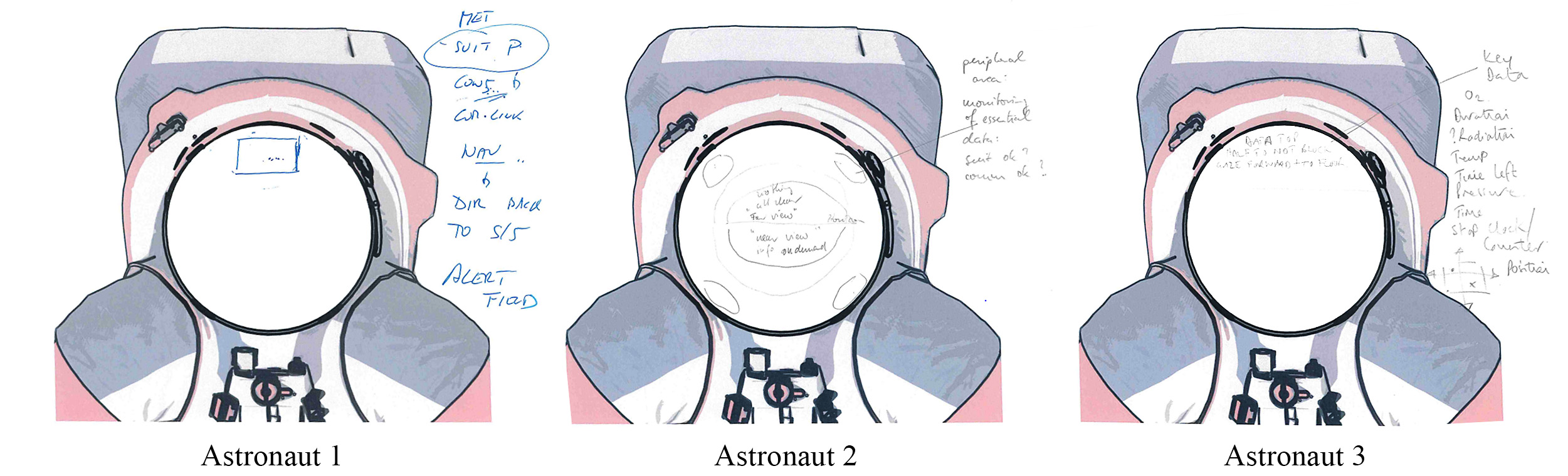}
\captionsetup{justification=centering}
\caption{Three conceptual sketches for HUD layouts, strategically positioning informational elements at the upper region of the HUD interface. This design choice is predicated on the absence of mission-critical visual cues in the lunar sky, thereby maximizing the utility of this visual space. Conversely, the lower field of view is deliberately left unobstructed to facilitate unimpeded visibility of the lunar terrain, thereby aiding astronauts in the identification and avoidance of potential navigational obstacles or interesting geological samples.}
\Description{The figure shows a series of conceptual sketches for HUD layouts, strategically positioning informational elements at the upper region of the HUD interface. This design choice is predicated on the absence of mission-critical visual cues in the lunar sky, thereby maximizing the utility of this visual space. Conversely, the lower field of view is deliberately left unobstructed to facilitate unimpeded visibility of the lunar terrain, thereby aiding astronauts in the identification and avoidance of potential navigational obstacles or interesting geological samples.}
\label{fig:figure3}

\end{figure*}

Going a step further, Astronaut 2 then divided the primary focus area into two halves along the lunar horizon: the "far view" positioned at the top, covering the user's lunar sky perspective, and the "near view" located at the bottom, encompassing the lunar ground (see Figure \ref{fig:figure3}). His rationale was that keeping track of the lunar sky would typically not be essential for astronauts on the Moon. Therefore, this section of the helmet view could be utilized for displaying robust navigational tools, such as maps. Conversely, he stressed the importance of maintaining maximum clarity in the near view, arguing it should only be used to present real-time visual cues relevant to the astronaut's current locus of action. 

Similar layout principles were largely echoed by other participants too. Leveraging his aviation expertise, Scientist 1 recommended placing vital signs indicators at the top of the user’s view rather than the sides, suggesting that maintaining clear lateral vision of the lunar environment would be more important than the vertical field of view. Scientist 7 argued the absence of weather phenomena on the Moon makes the upper half of the user’s view generally uneventful, rendering it a suitable location for HUD element display. Instructor 1 further pointed out that the dark lunar sky visible in the user’s upper view would offer better contrast, enhancing the legibility of digital information. 

Concerning the lower half of the helmet view, all participants recognized the importance of astronauts being able to watch their steps while navigating the lunar terrain. This had several implications for the HUD layout.

Most notably, it was seen as crucial that navigational cues did not force astronauts to divert their attention from the terrain they were traversing. Some, such as Scientist 4 and Instructor 1, criticized the compass interface in our VR simulation for requiring users to look up to read navigation instructions, potentially even causing “tunnel vision”. In contrast, Scientist 5 and Instructor 3 praised the AR pathway interface due to being projected directly onto the terrain, enabling astronauts to maintain focus while walking. 

In addition to being within the user's primary focus, it was considered vital for these navigational cues to not obstruct the view of the lunar terrain. Astronaut 3 and Engineer 3 commented on the risk of poorly designed navigational elements accidentally concealing objects like rocks, which could lead to tripping hazards. In response, a common recommendation was to keep these cues minimalistic (e.g., a thin dashed line) or semi-transparent. 

Providing supportive navigational information without blocking crucial objects in the user's field of view or overwhelming them with excessive content proved to be a delicate challenge. Participants were generally in agreement that the ideal HUD layout would vary depending on the situation. What might seem supportive in one instance could easily become obtrusive in another. 

Following this line of reasoning, Instructor 1 and Astronaut 2 both highlighted information overload as a primary challenge for future HUD solutions. In response, many participants brought up the need for context-sensitive design. Scientist 1 suggested adopting “task-specific displays”. Scientist 7 argued that some situations might even necessitate fully disabling a HUD:

\begin{quote}
\textbf{Scientist 7:} "There could be situations where you would need to have your full view clear. Like during vehicle ingress. So maybe some information would need to disappear. But again, the displayed information is not static. That's the beauty of a digital display, you can choose the amount of information.” 
\end{quote}

In line with Scientist 7’s suggestion, the majority of participants stressed the importance of adaptability for navigational HUDs. Astronaut 1 recommended that astronauts have the freedom to toggle specific navigational features on or off based on situational needs. Engineer 3 emphasized that the position of HUD elements ought to be adjustable too. Instructor 1 noted that spacesuits with built-in HUDs might be shared among multiple crew members, each with their unique preferences, further reinforcing the necessity for customization. 

In summary, it became clear that due to the dynamic and often unpredictable nature of EVA operations, the users’ needs with regards to their navigation HUDs would vary. To accommodate for this, some level of user control would be essential.

\subsection{Controlling a Lunar HUD}

Whilst our participants were in agreement about the necessity of giving astronauts control over their HUDs, there was some discord regarding the extent and nature of this control.

Engineer 5 pointed out that although customizability is important, it should have its limits. Astronauts should not, for example, have the ability to deactivate safety critical features or mute key communication channels. In addition, Engineer 7 noted that providing astronauts with an extensive range of customization options might prove overwhelming and add to the already stressful nature of EVA operations. He argued that all EVA interfaces need to strike a balance between "offering flexibility and providing reliable support”. 

In light of such arguments, participants generally leaned toward relatively simple control mechanisms. For instance, Instructor 4 proposed that navigational HUDs should offer a selection of preset configurations that the astronauts could easily switch between, rather than providing fine-grained control over every aspect of the HUD. 

Physical buttons were frequently brought up as the most straightforward means for enabling such control. However, it quickly became evident that they might not be well-suited for lunar EVA operations. Scientist 2 pointed out that buttons would need to be oversized to accommodate the bulky astronaut gloves. Instructor 4 agreed, reflecting on the downsides of button-based interfaces:

\begin{quote}
\textbf{Instructor 4:} "I would not like buttons. You will always do some work while on the Moon, you will be handling tools, you will have all sorts of stuff on yourself that can interfere with the physical buttons. Also, you're wearing an EVA suit with gloves that have probably the worst haptics you can possibly think about. The last thing you want is to rely on your hands for fine motor activities." 
\end{quote}

Scientist 6 also pointed out that physical buttons, like any other mechanical interface on the Moon, would be susceptible to contamination by regolith (moondust) and the risk of jamming. He speculated about circumventing this problem by installing a button-based interface on the interior of astronauts' gloves.

Nevertheless, there was a prevailing consensus that any viable control interface should ideally be hands-free. Voice-based interaction emerged as perhaps the most popular control solution among our participants. Astronaut 3 regarded it as the "easiest solution," while Instructor 1 argued that it would be "convenient" compared to alternative interfaces. Astronaut 2, on the other hand, took a more cautious stance, expressing concerns about interference with internal suit sounds and voice communication with the mission control center:

\begin{quote}
\textbf{Astronaut 2:} "Voice control would be okay, but in the helmet, you know, we have an airflow, so I'm not sure. Usually, we are hot-mic’d, so whenever I talk, the ground [mission control center] would hear it. So for voice control, we need to consider whether the overall operational concept would allow us to only talk inside the suit with ourselves using the voice control system and switch active radio comms on and off." 
\end{quote}

An alternative control solution that attracted some discussion involved the use of machine learning algorithms. Engineer 7, for example, suggested that AI systems could learn from an astronaut's training activities to anticipate their preferences and behaviors, allowing, in turn, for automatic contextual HUD adaptations without the need for manual input. Other participants disagreed, with Engineer 3 expressing concern about depriving users of control. 

Finally, an often debated topic was the notion of delegating control over one’s navigational HUD to a relevant mission control center on Earth. In this regard, there was a strong consensus among participants that control should remain in the hands of the astronauts. Scientist 7 pointed to their superior situational awareness. Instructor 1 and Astronaut 3 both argued the high communication latency and potential radio blackouts which would render such a remote control impractical. 

On the other hand, Engineer 5 noted that whilst astronaut autonomy is important, the mission control center should be free to edit certain content in the astronaut’s HUD, such as traverse paths or mission checklists, to keep the astronaut updated concerning developing situations. 

Discussions on the optimal allocation of responsibilities between astronauts and control centers sparked frequent reflections concerning the astronaut’s role within the broader mission architecture. As we shall explore in the following subsection, these considerations hold potentially important implications for navigational HUDs.

\subsection{Lunar HUDs in a Broader Operational Context}

The difficult communication conditions on the Moon, in combination with the limited situational awareness of mission control centers, led the majority of participants to argue that astronauts in the field ought to have the capacity to carry out navigational operations in an unsupervised manner. This should include the freedom to alter any predetermined course of action if the situation demands it. Scientist 1, for instance, argued that unexpected discoveries, such as rare mineral deposits, might be made during EVA missions, justifying a change of route. Instructor 1 added that navigational guidelines provided by HUDs might not always be accurate, making it essential for astronauts to rely on their own judgment as a primary resource.

Participants identified several ways in which navigational HUDs could cater to this need for astronaut autonomy. Scientist 7, for example, critiqued the detailed AR pathway in our virtual environment, warning against excessive overreliance. Should such an interface malfunction halfway through an EVA traverse, he reasoned, astronauts could be left stranded. Instead, he advocated delivering navigational instructions in a manner that promotes active exploration and understanding of the nearby environment.

Following a similar line of reasoning, Engineer 7 and Instructor 1 proposed a more flexible navigational interface in the form of a series of waypoints. This approach would support the astronaut’s general orientation while allowing them the freedom to select their own path from one waypoint to the next. Engineer 7 added that a fully predefined path would only be acceptable if accompanied by “very good operational justifications”. This viewpoint was echoed by Astronaut 3, who argued a HUD should explain the underlying reasons behind the navigational instructions it provides:

\begin{quote}
\textbf{Astronaut 3}: [commenting on the AR pathway interface] “I think it's a bit prescriptive. And there should be a particular reason for that. So what's the reason for it? I think if you're going through a minefield, then this would definitely be useful. But it would have to be very, very reliable. So it has to be meaningful. Why should I follow that? Why can I not just take a slightly different route?” 
\end{quote}

In other words, rather than simply instructing them where to go, navigational HUDs should provide the astronauts with meaningful explanations, enabling them to make their own informed decisions. 

Several participants believed that for astronauts to make such informed decisions independently, they would need access to a “big picture” overview, not just concerning their geographic situation, but likewise their progress with regards to the overall mission plan. Astronaut 2's comment below illustrates the importance of this information in guiding astronauts' navigational decisions:

\begin{quote}
\textbf{Astronaut 2:} "When you plan an EVA, you have 7 hours of oxygen. You want to know if you're in the scheduled area because you can't burn 50\% of your time on the first 10\% of the targets. This information needs to be available.” 
\end{quote}

Similarly, Instructor 4 argued astronauts ought to be able to see their traverse history, as this could prove invaluable in supporting their orientation or even guiding them back to safety in case of an emergency. 

It thus became evident that adequate navigational support would encompass a comprehensive set of information from which the astronauts could triangulate to determine their next course of action. Some participants, like Instructor 1, suggested combining multiple HUD interfaces, such as AR waypoints for short-range guidance and a minimap for a broader perspective. Others came full circle, arguing that handheld devices, or even paper notebooks, could complement the HUD with mission-critical meta-data. Nevertheless, all participants agreed that the integration of navigational HUDs and their content should be carefully coordinated within the broader human-machine ecosystem facilitating astronaut operations on the Moon.

\subsection{Quantitative Measurements}
 Herein we provide an overview of quantitative findings related to usability and perceived task load reported by our participants after interacting with our simulated analogue environment. Additionally, we reflect on perceived presence, route deviation, and task completion times. 
 The limitations stemming from this approach will be further elaborated in the Discussion section. 

\subsubsection{Presence}
Participants reported a notably high sense of presence within the virtual environment, with a mean (M) score of 5.23 and a standard deviation (\textit{SD}) of 1.15 on a scale from 1 (not at all) to 7 (very much). This confirms our VR testbed was immersive enough to lead our investigations.

\begin{table*}[htp]
    \centering
    \label{tab:SummaryTable}
    \begin{tabular}{lcccccccc}
        \toprule
        Condition & \multicolumn{2}{c}{Time (sec./min.)} & \multicolumn{2}{c}{Deviation (m)} & \multicolumn{2}{c}{NASA TLX} & \multicolumn{2}{c}{SUS Scores} \\
        \cmidrule(r){2-3} \cmidrule(r){4-5} \cmidrule(r){6-7} \cmidrule(r){8-9}
        & \textit{M} & \textit{SD} & \textit{M} & \textit{SD} & \textit{M} & \textit{SD} & \textit{M} & \textit{SD} \\
        \midrule
        Tablet   & 410 (6:50 min.) & 170 (2:50 min.) & 18.90 m & 7.43 m & 59.23 & 29.13 & 46.14 & 23.47 \\
        Minimap   & 273 (4:33 min.) & 114 (1:54 min.) & 15.85 m & 4.15 m & 27.68 & 15.62 & 77.50 & 15.39 \\
        AR Pathway & 273 (4:33 min.) & 72 (1:12 min.) & 10.49 m & 4.93 m & 18.09 & 13.60 & 85.23 & 13.52 \\
        Compass   & 440 (7:20 min.) & 100 (1:40 min.) & 14.41 m & 4.48 m & 36.09 & 20.51 & 69.32 & 17.61 \\    
        \bottomrule
    \end{tabular}
    
    \captionsetup{justification=centering}
    
    \caption{\textit{M} and \textit{SD} for the Time to Complete the Traverse Route, Deviation From the Recommended Traverse Route, NASA TLX Raw Scores, and System Usability Scale (SUS) by Condition.}

    \Description{Table 2 offers a comprehensive summary of quantitative metrics across the four experimental conditions— AR Pathway, Compass, Tablet, and Minimap. The table is organized into seven columns. The first column, labeled 'Condition,' identifies the four navigational aids under investigation. The subsequent columns are grouped into three pairs, each representing a different performance metric: Time to complete the traverse route, Distance from the recommended traverse route (in meters), NASA Task Load Index (TLX) Raw Scores, and System Usability Scale (SUS) Scores. Within each pair of columns, the first column presents the mean (\textit{M}) and the second column provides the standard deviation (\textit{SD}) for the respective metric.}
    
\end{table*}

\subsubsection{NASA TLX RAW Scores}
\begin{figure}[htp]
  \centering
  \includegraphics[width=0.5\textwidth]{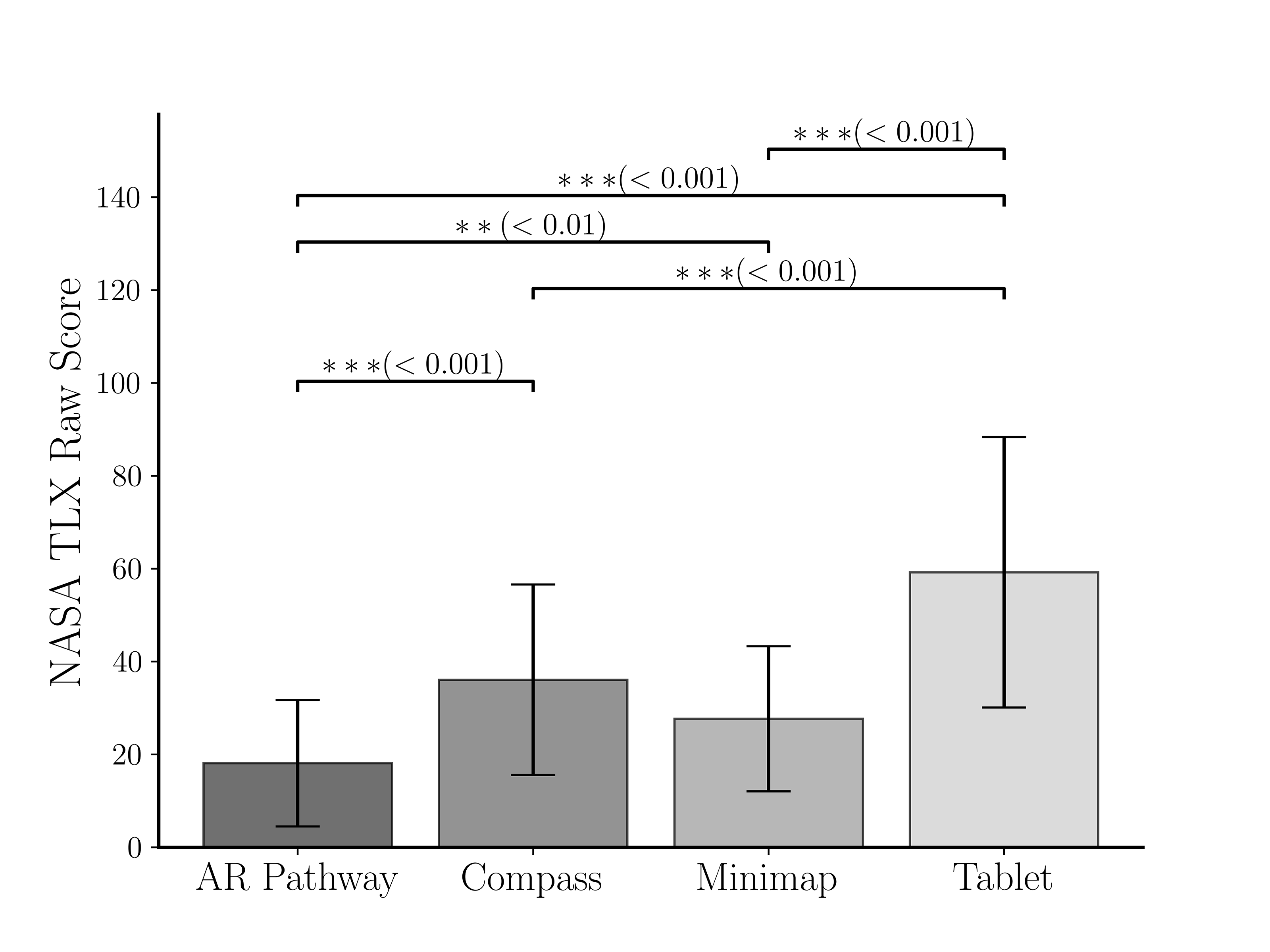}
  \caption{NASA TLX Raw Scores: Mean (\textit{M}), Standard Deviation (\textit{SD}), and significance levels for pairwise comparisons across experimental conditions.}
  \label{fig:NASATLX}
  \Description{Figure 5 presents a bar chart depicting the NASA Task Load Index (TLX) Raw Scores across various experimental conditions. Each bar represents the mean (\textit{M}) score for a specific condition, and error bars indicate the standard deviation (\textit{SD}) around the mean, providing a visual representation of the data's variability. Significance levels for pairwise comparisons between conditions are also annotated on the chart, offering insights into the statistical differences between the navigational aids under investigation}
\end{figure}

The results of the rmANOVA indicated a significant effect of the navigation feature on the raw scores of the NASA TLX \(F(3, 63) = 32.30, p < 0.001\) (see \autoref{fig:NASATLX}).
The AR Pathway condition was associated with significantly lower NASA TLX scores than the Compass condition, \(t(21) = 5.376, p < 0.001\) (corr.), as well as the Minimap condition, \(t(21) = -3.962, p = 0.004\) (corr.), and the Tablet condition, \(t(21) = -6.618, p < 0.001\) (corr.).
The Minimap condition had significantly lower raw NASA TLX scores than the Tablet condition, \(t(21) = -5.442, p < 0.001\) (corr.), and the Compass condition, \(t(21) = 2.833, p = 0.060\) (corr.).
Moreover, the Compass was associated with significantly lower NASA TLX scores than the Tablet condition, \(t(21) = -5.658, p < 0.001\) (corr.). The NASA TLX ratings thus further highlight our participants' belief that using a handheld tablet for navigational purposes is indeed associated with a higher overall workload compared to HUD-based solutions, therefore validating \textbf{H1}.  

\subsubsection{System Usability Scale (SUS) Scores}
\begin{figure}[htp]
  \centering
  \includegraphics[width=0.5\textwidth]{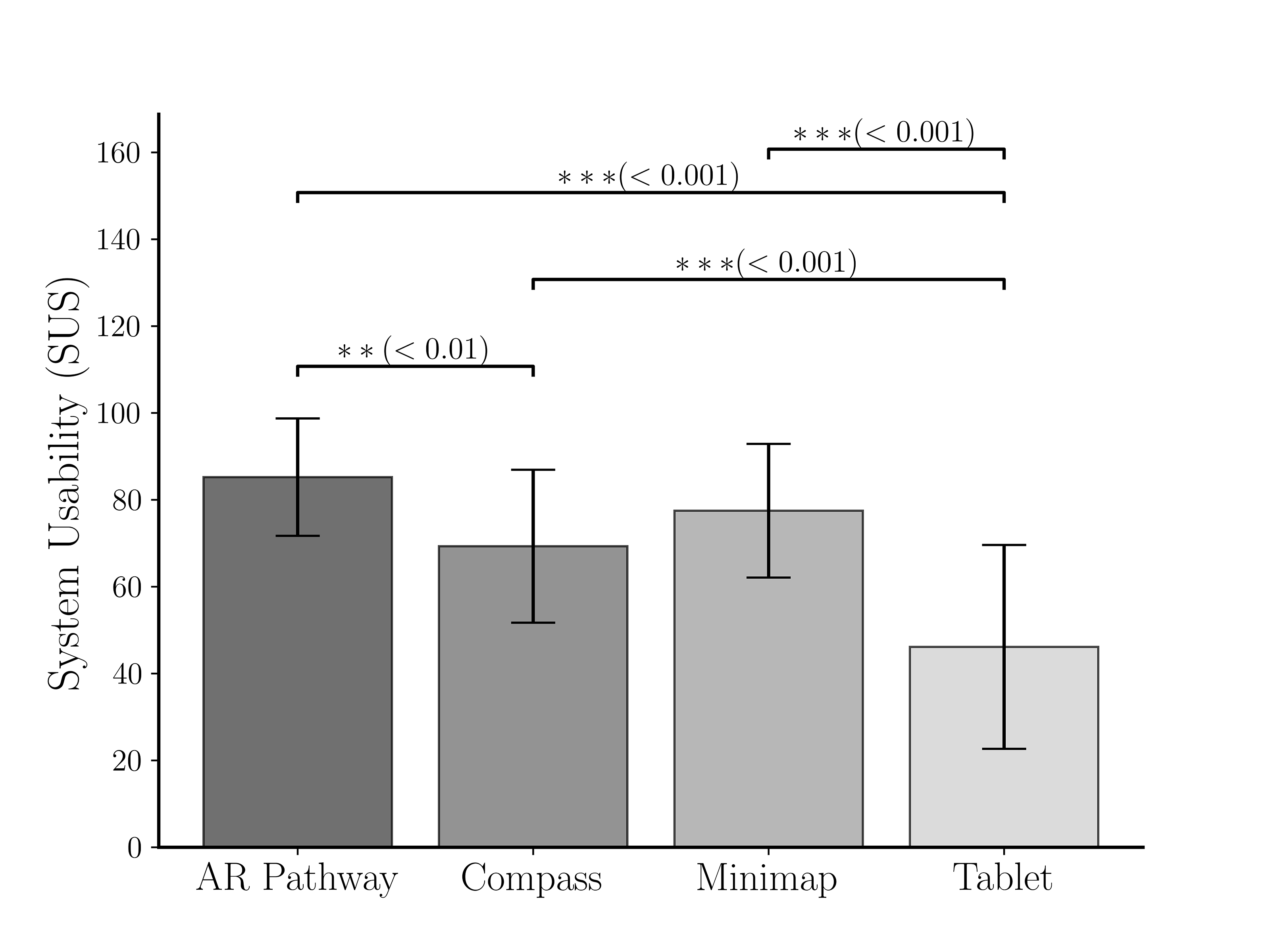}
  \caption{SUS Scores: Mean (\textit{M}), Standard Deviation (\textit{SD}), and significance levels for pairwise comparisons across experimental conditions.}
  \label{fig:SUSRating}
    \Description{Figure 6 presents a bar chart depicting the System Usability Ratings across various experimental conditions. Each bar represents the mean (\textit{M}) score for a specific condition, and error bars indicate the standard deviation (\textit{SD}) around the mean, providing a visual representation of the data's variability. Significance levels for pairwise comparisons between conditions are also annotated on the chart, offering insights into the statistical differences between the navigational aids under investigation}
\end{figure}

A significant effect of the navigation feature on the ratings of the SUS was observed, as evidenced by \(F(3, 63) = 28.131, p < 0.001\) (see \autoref{fig:SUSRating}). The AR Pathway was associated with significantly higher SUS scores than both the Compass, \(t(21) = -3.641, p = 0.009\) (corr.), and the Tablet, \(t(21) = 6.542, p < 0.001\) (corr.). Additionally, both the Compass, \(t(21) = 4.964, p < 0.001\) (corr.), and the Minimap, \(t(21) = -2.352, p = 0.171\) (corr.), were rated as having significantly higher usability than the Tablet. No significant differences were found between the Compass and the Minimap, \(t(21) = -2.352, p = 0.171\) (corr.), and between the AR Pathway and Minimap, \(t(21) = 2.865, p = 0.056\) (corr.).

Overall, a score above 68 is considered as indicating above-average usability, whereas a score below 68 is considered as below average \cite{lewis2018item}. Given the mean scores of the four conditions, therefore, only the tablet would be considered as having a below-average SUS rating, whereas the compass, the minimap, and the AR pathway all had above-average usability ratings. Therefore, the SUS scores additionally support the findings of the qualitative analysis, indicating the superiority of the HUD elements in terms of usability, thus validating \textbf{H2}. 

\subsubsection{Deviation From the Recommended Traverse Route}

\begin{figure}[htp]
  \centering
  \includegraphics[width=0.5\textwidth]{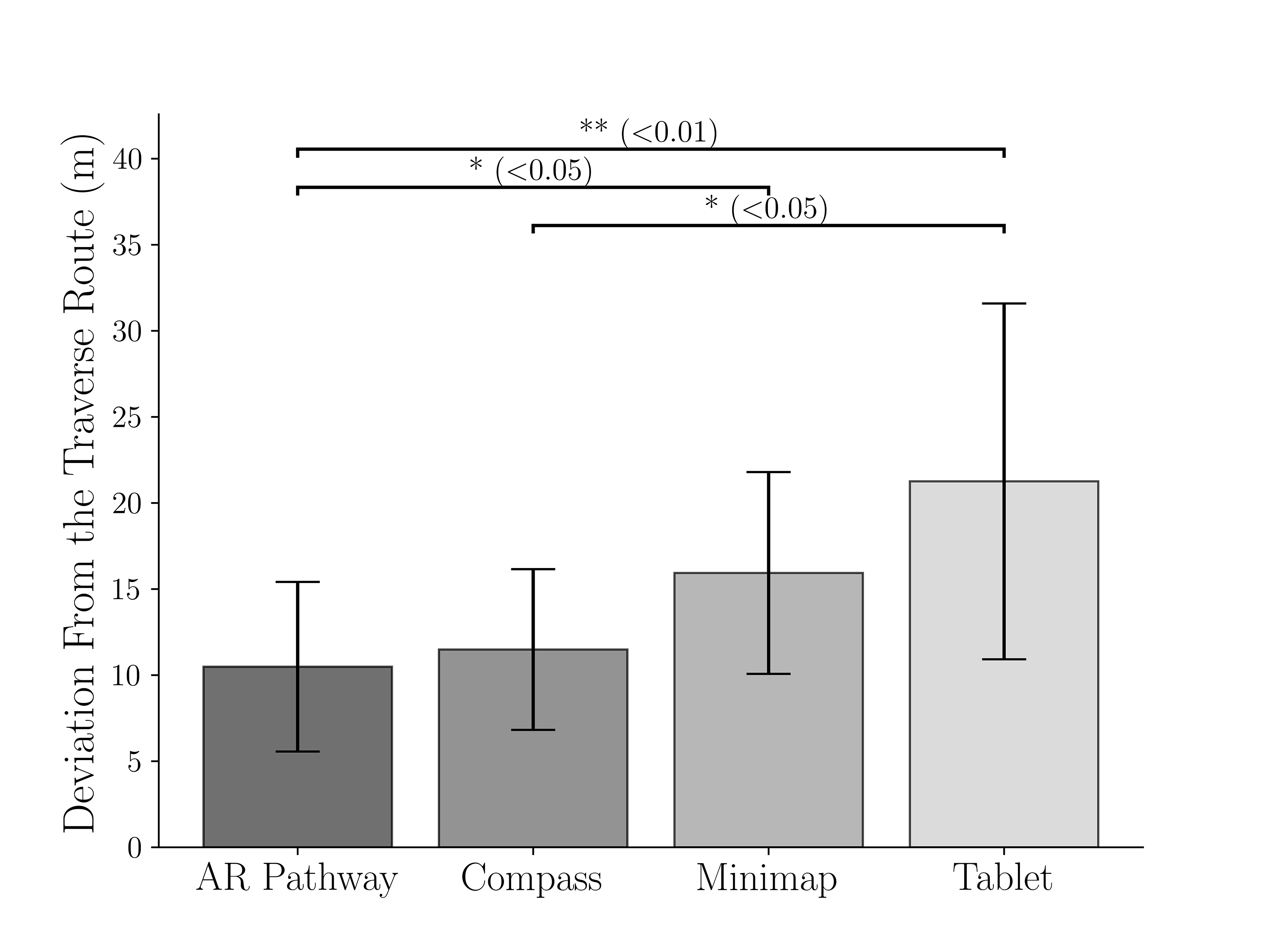}
  \caption{Deviation from the  Recommended Traverse Route: Mean (\textit{M}), Standard Deviation (\textit{SD}), and significance levels for pairwise comparisons across experimental conditions.}
  \label{fig:Distance}
  \Description{Figure 7 presents a bar chart depicting the Distance from the Recommended Traverse Route across various experimental conditions. Each bar represents the mean (\textit{M}) score for a specific condition, and error bars indicate the standard deviation (\textit{SD}) around the mean, providing a visual representation of the data's variability. Significance levels for pairwise comparisons between conditions are also annotated on the chart, offering insights into the statistical differences between the navigational aids under investigation}
\end{figure}

The results of the rmANOVA indicated a significant effect of the navigation feature on the Deviation from the Recommended Traverse Route \((F(3, 48) = 9.0247, p < 0.0001)\) (see Figure \ref{fig:Distance}). Post-hoc pairwise comparisons revealed that the AR Pathway condition emerged as particularly effective, significantly outperforming the Tablet condition, \(t(16) = -3.8780, p < 0.01\) (corr.), and the Minimap condition, \(t(16) = -3.1596, p < 0.05\) (corr.). The Compass condition did not significantly differ from the AR Pathway condition, \(t(16) = 0.8427, p  = 1\) (corr.), however, it significantly outperformed the Tablet condition, \(t(16) = -3.3593, p < 0.05\) (corr.). It also showed a trend of better performance compared to the Minimap condition, although this was not statistically significant, \(t(16) = -2.8796, p  = 0.0654\) (corr.). No significant effect was visible when comparing the  Minimap condition against the Tablet condition  \(t(16) = -1.7524, p  = 0.5931\) (corr.). Therefore, it appears that the participants objectively deviated even more from the route when using the tablet compared to most HUD-based conditions and especially AR guidance cues.

\begin{figure}[h]
  \centering
  \includegraphics[width=0.5\textwidth]{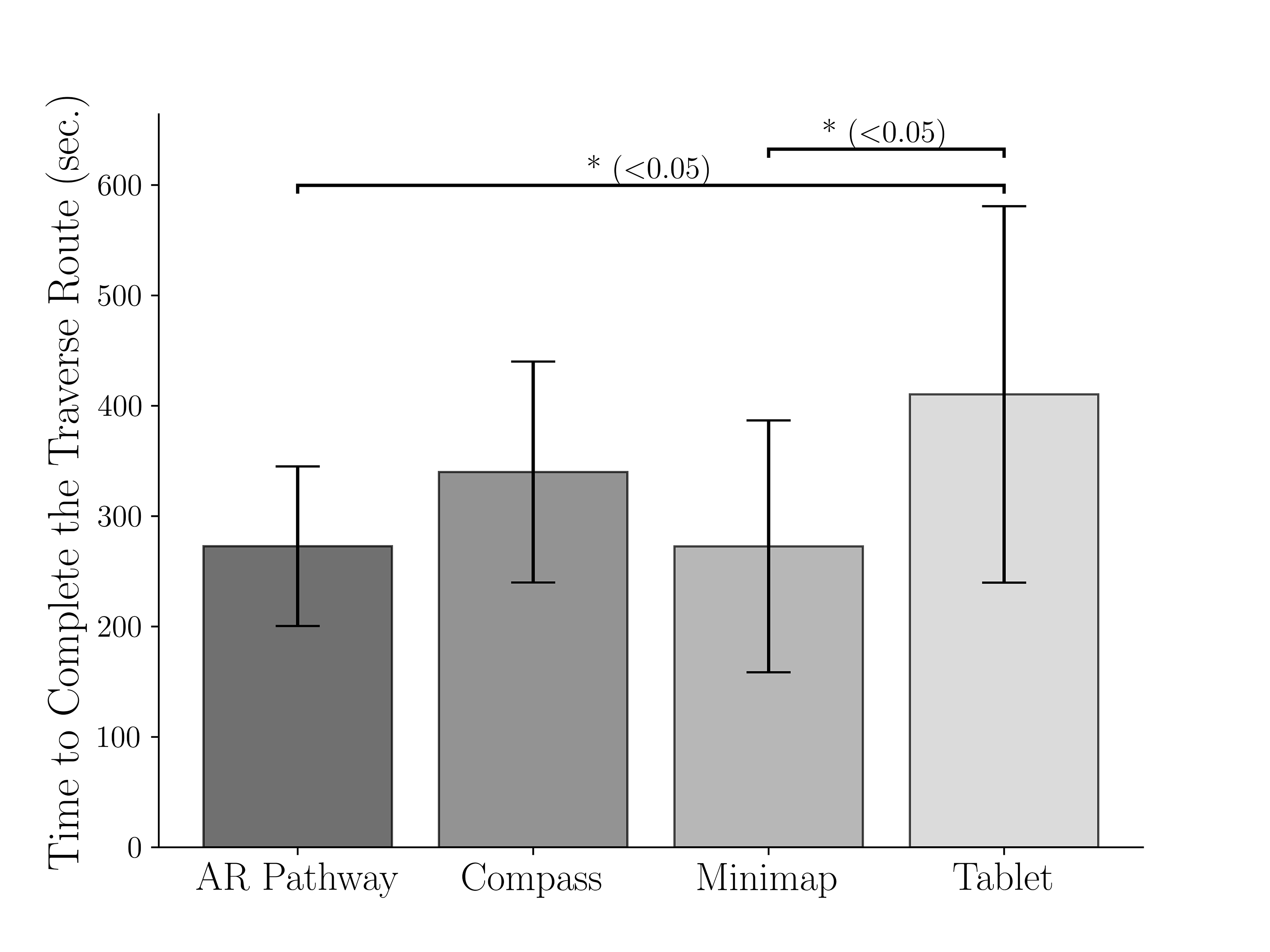}
  \caption{Time to Complete the Traverse Route: Mean (\textit{M}), Standard Deviation (\textit{SD}), and significance levels for pairwise comparisons across experimental conditions.}
  \label{fig:TTC}
  \Description{Figure 8 presents a bar chart depicting the time needed to complete the route across various experimental conditions. Each bar represents the mean (\textit{M}) score for a specific condition, and error bars indicate the standard deviation (\textit{SD}) around the mean, providing a visual representation of the data's variability. Significance levels for pairwise comparisons between conditions are also annotated on the chart, offering insights into the statistical differences between the navigational aids under investigation}
\end{figure}

\subsubsection{Time to Complete the Traverse Route}

When analyzing the \textit{Time to Complete the Traverse Route}, our study revealed significant differences between Conditions ($F(3, 48) = 7.637, p < 0.0001$) (see Figure \ref{fig:TTC}). Post-hoc pairwise comparisons revealed that participants completing the AR Pathway condition were significantly faster than in the Tablet condition \(t(16) = -3.6452, p < 0.01\) (corr.). In contrast, the Minimap condition did not exhibit a significant difference when compared to the AR Pathway condition \(t(16) = 0.0043, p = 1\) (corr.), but it significantly outperformed the Tablet condition \(t(16) = -3.1069, p < 0.05\) (corr.). There was no significant differences between Compass and AR pathway \(t(16) = 2.5132, p = 0.1283\) (corr.) , Minimap \(t(16) = 2.5300, p = 0.1337\) (corr.) and Tablet conditions \(t(16) = -1.8287, p = 0.5170\) (corr.). 

\section{Discussion}

Humanity’s return to the Moon entails a multitude of human and technical challenges. This paper explores the issue of lunar surface navigation and the design of HUD interfaces to tackle it. By offering pertinent information directly in an astronaut's helmet, HUDs have already shown promise in aiding astronauts during EVAs in Low Earth Orbit \cite{coan_exploration_2020, cardenas2021aaron}. However, the design of such solutions for the unique lunar context has historically faced limitations due to the frequent necessity for experimental deployments in analogue environments. 

Drawing on the Human Systems Exploration method, we have immersed 25 domain experts into a representative VR-based lunar environment and interactively evaluated four different navigational interfaces. Our findings demonstrate the viability of such simulated analogues in facilitating assessments of early-stage HUD design concepts. The trends identified through our approach align broadly with findings from traditional real-world studies, suggesting a promising avenue for Human-Computer Interaction (HCI) design in future lunar systems. Below, we elaborate on this by reflecting on key design challenges and potential solutions brought up by our participants.

\subsection{Towards a Lunar HUD}

\subsubsection{Mitigating Challenges in the Lunar Environment}

Much like the Apollo astronauts before them \cite{apollo11_1969}, our participants found the lunar lighting conditions challenging. Based on qualitative reflections of the participants and supporting quantitative data (NASA TLX, SUS, performance measures) our study suggests that indeed HUDs and in particular spatial 3D elements could prove instrumental in supporting navigation in extreme lighting conditions, especially in comparison to suit-mounted or heads- down navigational tools. Such spatial guidance information should furthermore be enhanced with other depth cues (e.g., shadowing, texture gradients or checkerboard pattern).

Nevertheless, 2D representations (e.g. colored outlines) appeared to retain some value for understanding object positioning, a finding supported by the work of Bruno et al. \cite{bruno1988minimodularity}. A key navigational feature in this vein brought up by our participants was the highlighting of obstacles, such as rocks or other tripping hazards. In this context, relevant research demonstrated that world-locked cues utilizing object tracking to overlay obstacles provide superior efficiency compared to directional cues during navigational tasks \cite{fox2023using}.

Therefore, our results suggest that guidelines established in the aerospace domain, specifying that so-called conformal guidance cues and symbology, where spatial overlays are matched to the real world and stay aligned even when moving, can be extended to the specific lunar context \cite{wickens1995object,wickens2003aviation,mccann1996scene,long1994conformal}.

Furthermore, it seems reasonable to assume that such findings would likewise apply to the highlighting of interesting terrain features or anomalies for exploration, such as unusual geological formations \cite{ohgushi2020road}.

Facing poor terrain legibility in shadowed areas, participants commonly expressed the need for elevation data on minimaps, echoing the findings of St. John et al., who demonstrated the utility of 3D maps in representing terrain depth \cite{John2001TheUO}.

Our participants also emphasized the need for colors that offer strong contrast against the lunar environment whilst matching the already established color standards.
To mitigate the lunar environment, future HUDs should feature:

\begin{itemize}
    \item \textbf{Spatial conformal cues}, such as AR markers projected onto the lunar surface to enhance terrain legibility. Vertical cues, such as flag objects, would further enhance distance estimation.
    \item \textbf{Conformal 2D cues}, such as "colored circles", object outlines or a terrain overlay to indicate no-go zones and other safety hazards.
\end{itemize}

\subsubsection{Avoiding Information Overload}

Several participants expressed concerns regarding excessive amounts of information in their HUDs. The importance of mitigating information overload has been highlighted by prior work, with various “declutter capabilities” being discussed that would allow the user to reduce the non-essential visual information displayed in their HUD \cite{ververs1998head,fadden2001pathway,fischer1980cognitive}. 

Fadden et al., found that the risk of clutter is generally lower with spatial navigation cues, but likewise warned against spatial cues that are too dominant \cite{fadden2001pathway}. Similar concerns were reported by our participants, who suggested that the spatial AR pathway constituted a potential safety hazard by occluding information on the ground (e.g. boulders) while causing the participants to pay too much attention to the pathway at the expense of the rest of the scenery.

Importantly, research has shown that providing excessive information could indeed be especially detrimental during EVAs that are generally associated with a high degree of workload, as the general susceptibility to attentional tunneling increases with higher workload  \cite{weintraub1992book, prinzel2004head, stuart2001head}. To counteract such challenges, lunar HUDs should support: \\

\begin{itemize}
   \item \textbf{Offloading of nonessential information}, such as to a suit-mounted tablet or other forms of head-down displays. 
  \item \textbf{Adopt a layout that aligns with the lunar environment}, such as by concentrating pertinent information at the top of the user’s view against the lunar sky, while keeping the surrounding terrain unobstructed.
\end{itemize}

\subsubsection{Supporting Astronaut Autonomy}
Future lunar ground crews will face challenges related to low bandwidth and poor latency issues, which may hinder efficient support from the mission control center during critical situations. It will thus be essential for astronauts to have the capability to operate independently. 

Our study suggests this does not necessarily translate into the need for more robust navigational solutions. Whilst strong navigational cues, such as the spatial AR pathway, achieved superior workload ratings, it still faced criticism for being overly prescriptive and fostering overreliance. Instead, our participants expressed the desire for opportunistic navigation, such as by engaging in exploratory behavior. A similar phenomenon has also been frequently observed in the military context. The concept of "HUD-cripple," originally coined by Navy aviators, denotes an over-reliance on HMDs to the detriment of traditional skill sets \cite{lindberg2020augmented}. 
Other studies went a step further, arguing that strong navigational cues could lead to attention tunneling \cite{fadden2001pathway}. 

In this vein of inquiry, prior research has differentiated between command and situation guidance cues \cite{wilson2002comparing}. While the former provides the user with explicit instructions for what needs to be done, the latter aims to support the users in making their own informed decisions. 

Our study suggests that both, local and global perspectives are needed to allow for such a level of informed user autonomy. As expressed by our participants, astronauts need to navigate their immediate surroundings, as well as to make navigational decisions based on macroscopic considerations, such as the overall mission plan, traverse history, or EVA elapsed time. 

This dual-level approach, such as spatial navigation cues for local information and a top-down map featuring a broader contextual view, has already been shown to improve situational awareness in aviation settings \cite{mccann1996advanced, andre1998field}, whilst our study additionally suggests a promising avenue for adaptation to astronaut HUD systems. In light of this, lunar HUD systems should: 
\\

\begin{itemize}
    \item \textbf{Offer situational guidance}; such as overview maps or adaptive waypoints flexibly responding to the astronauts own actions.

    \item \textbf{Support both local and global orientation} by offering astronauts information not only about their immediate surroundings but also providing a broader perspective on mission context, including overarching goals and potential points of interest.
\end{itemize}

\subsubsection{Making HUDs Explainable}
In a closely related theme, our study suggests that a navigational system conducive to user’s autonomy would need to have the capacity to explain the underlying factors behind navigation instructions, thus enabling the user to make an informed independent decision. A major topic of interest in HCI and Artificial Intelligence domains, explainability has been described as the users' ability to predict, control, and understand the behavior of a technical system \cite{linardatos2020explainable}. 

To advance explainability in AR systems, Xu et al. introduced the XAIR framework, which provides guidelines on the timing, content, and modality of explanations for AI systems embedded in everyday AR solutions \cite{xu2023xair}. 

An additional solution may be the incorporation and visualization of a system's confidence level, allowing its users to gauge the computational accountability of their HUD system \cite{schmitz2022fast}, which could also potentially counteract the attention tunneling effect that has been observed when guidance cues are reliable \cite{yeh2003head}. 

Our study suggests that granting astronauts a high degree of autonomy, while concurrently providing them with detailed explanations of safety constraints, likely enhances their adherence to these guidelines. Such constraints are not seen as limitations but as essential parameters that are meaningful to the astronauts. 

We thus found that by framing these constraints within the context of mission safety, astronauts are more likely to comprehend and adhere to these parameters. This approach is particularly salient for the success of extravehicular activities and long-term missions, where understanding and following safety guidelines are critical. A HUD should thus provide: 

\begin{itemize}
\item \textbf{Navigational instructions explainable via mission- critical considerations}, such as safety of the astronaut, or projected EVA time constraints.  
\end{itemize}

\subsubsection{Unobtrusive Interaction with HUDs}

Our study highlights the importance of control mechanisms enabling astronauts to customize their HUD content and layout based on situational needs.

Conventional buttons were deemed cumbersome and risky, echoing findings from the field of Urban Warfare Augmented Reality (UWAR) \cite{argenta2010graphical}. Gesture-based controls, integrated into astronaut gloves, emerged as one viable alternative. This aligns with Lee et al.'s work on gesture-based astronaut gloves for drone control \cite{lee2020astronaut}. However, the utility of gestures is questionable during EVAs when astronauts are engaged in tasks that occupy their hands.

Similar to, for instance, Rometsch et al. \cite{rometsch2022design}, our participants were likewise positively inclined towards voice interaction as the main interaction modality for astronaut HUD control. While most participants preferred such interactions, others cautioned against reliability issues and potential interference with mission control communication. 

Eye-tracking solutions may likewise be worthy of exploration, however, this would likely limit the astronaut’s ability to scan the environment to some degree, whilst reliable confirmation options would also need to be developed. 

Considering the distinct preferences voiced by our participants, it is thus likely that a viable lunar HUD interface should: 

\begin{itemize}
    \item \textbf{Integrate multiple interaction modalities:} such as by combining voice commands with buttons or gesture-based interaction. This approach would enable astronauts to access vital HUD functionalities through different means, adapting to their situation.
\end{itemize}
\subsection{The Simulated Analogue Environment}
Apart from being costly and logistically demanding, real-world deployments and evaluations of lunar prototypes have attracted criticism for underexposing key environmental elements, including the unique lighting conditions on the lunar south pole \cite{nilsson_using_2022}. Our alternative approach utilizes VR as a testbed to immerse participants in a realistic scenario and elicit context-specific comments and insights. Whilst a similar approach has been utilized in other domains \cite{kuliga2015virtual}, it has not yet been applied to the assessment of HUD solutions for lunar surface EVAs. Yet, numerous studies suggest that VR applications can significantly improve the identification of user requirements and needs during an engineering design process \cite{chang2022influence, hubenschmid2022relive, lee2019design, jetter2020vr}.  

Our study produced findings that closely align with previous research (see section \ref{interfaces}), highlighting, for instance, the significant advantages of Augmented Reality Heads-Up Displays (AR HUDs) over traditional navigational tools. The numerous insights shared by participants during their immersion in the VR environment enabled us to explore such principles and articulate design guidelines tailored to the unique conditions of the lunar surface.

Understanding user needs is key to developing complex systems such as HUD interfaces for lunar EVAs  \cite{iso199913407}. Given that significant investments of resources can be expected to be required for the development of such systems, it is important that a good foundation is laid for this work, and the potential use cases, challenges, and requirements are known before the system is designed.

A number of such attempts have recently taken place in the human spaceflight domain. Drawing on literature analysis and brainstorming sessions, for instance, De Medeiros et al. \cite{de_medeiros_categorisation_2022} formulated a categorization of applications for AR in human lunar exploration. Such categorizations can be used to relate individual findings from separate studies that might focus on or be biased towards specific scenarios or use cases to a wider framework in order to gain a broad insight into the potential value of AR systems for lunar EVAs.

However, the detailed insights provided by participants of our study also highlight a limitation of such categorizations, namely that in generalizing use cases, some specificity is lost. This illustrates the value of detailed and context-specific insights, emerging from the interplay of representative environments and prospective users. 

In summary, a wider utilization of VR-based simulated analogues in the human spaceflight domain could have a transformative impact. Beyond simply reducing some of the associated costs, the highly accessible nature of VR makes it well positioned to help open up relevant activities and discussions to wider audiences. By enabling a greater number of stakeholders with different professional backgrounds to contribute ideas and help shape our future beyond Earth, VR thus exemplifies a disruptive technology that has the potential to foster a more inclusive, innovative, and dynamic space industry.

\subsection{Limitations and Future Work}

Our decision to conduct user studies in a simulated lunar analogue comes with important limitations. Chiefly, the absence of physical stimulation, such as astronaut suit constraints or the sensation of lunar gravity, has likely affected our findings to some degree by underexposing or trivializing certain aspects of lunar navigation. 

For instance, wearing an astronaut suit in lunar gravity shifts the astronaut's center of mass significantly \cite{apollo11_1969}, an effect that could not be replicated through our VR setup.

Participants were likewise embodied in the virtual environment using a torso and pair of gloves, rather than a fully rendered avatar. It is worth noting, however, that the xEMU suit designed for future lunar missions obstructs the astronaut’s view of their lower body \cite{NASA2019}. Prior research also suggests that a full virtual body does not impact relevant skills, such as distance estimation, during  VR-based navigation tasks \cite{mcmanus2011influence}. Nevertheless, the exact impact of whole-body avatars on navigation in VR remains under-explored.

Similarly, our virtual testbed was used under Earth's gravity conditions. Consequently, the astronaut's hopping motions typical of Apollo EVAs were not re-enacted. While we could have visually approximated these movements, doing so might have increased the risk of motion sickness due to the disparity between visual and physical movements \cite{hettinger1992visually}. 

Given such limitations, the (self-reported) measurements collected in this study (e.g. usability and workload) along with performance measures (completion time) ought to be interpreted with caution. We would argue the expert knowledge and qualitative feedback of our participants did compensate for this lack of fidelity to some degree. Osterlund et al., for instance, found that a purely audiovisual simulation of aerospace maintenance operations in protective suits could elicit valid and actionable feedback from subject matter experts \cite{osterlund2012virtual}. Nevertheless, we would advise our findings are primarily seen as preliminary indicators or starting points for further, more detailed, investigations. More work is needed to corroborate and validate these initial findings. In particular, we would propose the following research directions: 
\begin{itemize}
\item Future work could investigate the use of passive haptic feedback \cite{muender2022haptic} in the form of astronaut glove mockups in association with set props, thereby more accurately reenacting the haptic properties associated with astronaut’s manual operations, such as the use of a tablet. Such a setup could either be tracked as a virtual representation or be used through a mixed-reality configuration.

\item Similarly, movement constraints imposed by the full astronaut suit are likewise in need of greater investigation. VR controller-based interaction should thus be replaced by more natural and advanced hand and leg tracking technologies. Accurately replicating the effects of wearing an astronaut helmet with correct Sun reflections is likewise in need of exploration.

\item It might also be beneficial for future studies to assess the potential use of gravity off-load systems in combination with VR. Such an interface could accurately simulate lunar gravity conditions both physically and visually during movements, allowing for a more accurate simulation of lunar gravity effects on locomotion.

\item Apart from tackling the problem areas associated with analogues simulated in VR, as elaborated through this section, we would suggest future work focuses on evaluating prospective lunar HUD solutions in a real-world setting while factoring in some of the relevant physical constraints.

   \item More work is needed to better understand the potential use of sounds to decrease the amount of visual information received by astronauts during lunar EVA. Identifying types of navigation information that can be replaced or enhanced by 3D audio cues, for instance, would be beneficial. Splitting navigational information into auditory and visual information could be especially helpful in high workload situations, as attentional resources (audio vs. visual) can be shared at the same time, reducing the risk of attentional tunneling and reduced performance (Wickens' Multiple Resource Theory, \cite{prinzel2004head}).

   \item We expect the interface revisions outlined above to help increase the overall simulation fidelity, likely resulting in enhanced behavioral validity of user’s embarking on simulated missions in VR testbeds. Nevertheless, the absence of these revisions in our study did not preclude our expert participants from bringing up numerous relevant reflections concerning future human operations on the Moon. This underscores the viability of VR as an enabler of \textit{Research through Design} (RtD) practices in aerospace engineering - a field whose resource-intensive nature has historically marginalized the use of such approaches \cite{nilsson2023using}. As another productive avenue for future research, we would therefore suggest corroborating and expanding on these findings by deploying additional design artifacts in VR-based testbeds with the explicit aim of furthering our understanding of issues surrounding human lunar exploration.

\end{itemize}

\section{Conclusion}
This paper has explored the design of navigational HUD interfaces for prospective human operations on the Moon. While HUDs have demonstrated promise in enhancing the safety and performance of astronauts, their design has often been linked with intricate analogue studies, limiting the widespread application of HCI design approaches. In search of an alternative approach, we have immersed experienced human spaceflight experts in a simulated analogue environment using Virtual Reality. This method produced findings comparable with real-world studies, corroborating several crucial advantages of HUDs over traditional navigation tools, such as improved usability and reduced cognitive load.
Drawing on our findings, we have identified key challenges and requirements related to the design of HUD technologies in the context of future lunar EVAs. On these grounds, we have formulated a set of practical design guidelines, including strategies for tackling harsh lighting conditions, mitigating the risk of information overload, and supporting astronaut’s autonomy. In exposing these challenges and making them accessible to HCI practice, this paper seeks to lay the foundation for future work. Fostering novel accessible methods for rapid human-centered design and evaluation will be crucial for the development of reliable, secure, and efficient navigational technologies to support upcoming missions to the Moon as we enter a new era of lunar exploration.

\begin{acks} 
We would like to thank Samantha Cristoforetti, Matthias Maurer, and John McFall for their participation in our study! Moreover, we extend our sincere thanks to the domain experts at the European Astronaut Centre (ESA EAC) for their support and insights.
\end{acks}
\balance

\bibliographystyle{ACM-Reference-Format}
\bibliography{bibfile}

\end{document}